\documentclass[12pt,preprint]{aastex}
\usepackage{lscape}
\slugcomment{draft of \today}
\begin{document}
\newcommand\etal{{\it et al.~}}
\newcommand\cf{{\it cf.~}}
\newcommand\eg{{e.g.,~}}
\newcommand\ie{{i.e.,~}}
\newcommand\mug{{$\mu$G}}
\newcommand\control{{Control Model}}
\newcommand\injection{{Injection Model}}
\newcommand\cooling{{Cooling Model}}
\newcommand\spc{~}
\title{Synthetic Observations of Simulated Radio Galaxies I: 
Radio and X-ray Analysis}
\author{  I. L. Tregillis       \altaffilmark{1},
          T. W. Jones		\altaffilmark{2},
          Dongsu Ryu            \altaffilmark{3}} 
\altaffiltext{1}{Applied Physics Division, MS B259,
Los Alamos National Laboratory, Los Alamos, NM, 87545; iant@lanl.gov}
\altaffiltext{2}{School of Physics and Astronomy, University of Minnesota,
Minneapolis, MN 55455; twj@msi.umn.edu}
\altaffiltext{3}{Department of Astronomy \& Space Science, Chungnam National
University, Daejeon, 305-764 Korea; ryu@canopus.chungnam.ac.kr}
%
\begin{abstract}
We present an extensive synthetic observational analysis of
numerically-simulated radio galaxies designed to explore the
effectiveness of conventional observational analyses at recovering
physical source properties. These are the first numerical
simulations with sufficient physical detail to allow such a study.
The present paper focuses on extraction of magnetic field
properties from nonthermal intensity information.
Synchrotron and inverse-Compton intensities were effective in
providing meaningful information about distributions and 
strengths of magnetic fields, although considerable care
was called for in quantitative usage of the information.
Correlations between radio and X-ray surface brightness
correctly revealed useful dynamical relationships between 
particles and fields, for example. Magnetic field strength 
estimates derived from the ratio of X-ray to radio
intensity were mostly within about a factor of two of the
RMS field strength along a given line of sight. When
emissions along a given line of sight were dominated by
regions close to the minimum energy/equipartition condition,
the field strengths derived from the standard power-law-spectrum 
minimum energy calculation were also reasonably close to actual field 
strengths, except when spectral aging was evident. Otherwise,
biases in the minimum-energy magnetic field estimation
mirrored actual differences from equipartition.
The ratio of the inverse-Compton-estimated magnetic field to
the minimum-energy magnetic field provided a rough
measure of the actual total energy in particles and
fields in most instances, although this measure was
accurate within only about an order of magnitude.  
This may provide a practical limit to the accuracy with which one
may be able to establish the internal energy density or pressure
of optically thin synchrotron sources.

\end{abstract}

\section{INTRODUCTION}	\label{bic:intro}
The synchrotron emission from extragalactic radio
sources is a function of both the local magnetic fields and the
relativistic particle populations residing within.  These two
components are important to the energy budget of such objects. So,
pinning down their relative contributions is crucial to understanding
their overall behavior.  Unfortunately, the optically-thin synchrotron
emission alone cannot be used to  extract the individual particle and field 
components. However, it has long been known that in principle
radio synchrotron observations can be combined with X-ray observations of
inverse-Compton scattered cosmic microwave background photons
(hereafter IC/3K) to extract
information about particles and fields in emission
regions \citep[\eg][]{Harris74,Cooke78,Fabbiano79,Harris79}. 
The advent of the Chandra and XMM-Newton observatories has 
made this kind of analysis possible for a large number
of classes and objects.

X-ray emission has now been detected in the jets, hotspots, and lobes of
numerous radio galaxies and attributed to a host of different physical
processes ranging from synchrotron emission (\eg \citet{Wilson01})
to inverse Compton scattered emission off of one or more of
several photon fields. 
\citet{Brunetti99} and \citet{Setti02} reported the detection of X-ray 
emission in the lobes of 3C 219 and in 3C 215 and 3C 334 respectively,
that possibly arises from inverse-Compton
scattering of IR photons from the quasar nucleus.  
IC/3K lobe emission has been reported in several radio galaxies,
including 
Fornax A \citep{Kaneda95,Feigelson95}, 
Centaurus B \citep{Tashiro98},  
Abell85 0038-096 \citep{Bagchi98},
3C 295 \citep{Brunetti01},
3C 330 \citep{Hardcastle02a},
and
3C 263 \citep{Crawford03}.
Synchrotron self-Compton (SSC) emission
was reported in the hotspots of several powerful galaxies, including 
Cygnus A \citep{Harris94}, 
3C 295 \citep{Harris00}, 
3C 273 \citep{Roser00}, 
3C 123 \citep{Hardcastle01},
3C 207 \citep{Brunetti02}, 
and
3C 263 and 3C 330 \citep{Hardcastle02a}. 
Nonthermal X-rays of uncertain origin have recently 
been detected in the hotspots of 3C 280 and 3C 254 \citep{Donahue03}.

When combined with radio observations, X-ray detections allow one to
infer the magnetic field strength in an emitting region.
Yet, in practice the derived fields vary significantly, depending on the
assumed mechanism for X-ray emission.  So, interpretation 
is often uncertain. Indeed, closeness to the
equipartition value is sometimes used as a validation criterion for
a magnetic field estimated by any other means.  That seems
unacceptable, since there is
no convincing theoretical argument for equipartition between radio-emitting
electrons and magnetic fields, and empirical evidence argues
that not all sources are in equipartition (\eg
Centaurus B \citep{Tashiro98}, PKS 0637-752
\citep{Schwartz00}).  

There are obvious complications with all of the field measures. For example,
deconvolution of the particle and field 
information also requires assumptions about the particle and field
filling factors, as these values are 
impossible to extract from observations. For simplicity, a uniform
magnetic field distributed through the emitting region is customarily
assumed.  As explained below, caution is needed,
particularly when assuming magnetic field
uniformity.  It is important to be aware that the radio and X-ray
emissions may be dominated by physically different regions in the
source, and so one must be careful that the same
particle population is sampled by the radio and X-ray
observations.  
This paper is based on the expectation that synthetic observations of 
numerically simulated radio
galaxies may help us understand these issues in a way that makes minimal
use of convenient, simplifying mathematical assumptions, while
retaining the benefit of complete knowledge of the actual
physical conditions being observed.

\citet{Tregillis01a} (Hereafter TJR01) recently carried out three
dimensional time dependent MHD simulations of radio-jet flows that included 
nonthermal relativistic electron transport in space and momentum,
enabling them to create 
the first synthetic radio observations from 
nonthermal electron distributions that were
consistently evolved within the plasma flows.
Those synthetic observations were used in conjunction with a detailed
dynamical analysis to study how the dynamical and
nonthermal particle transport processes lead to observable radio
synchrotron surface
brightness and spectral index patterns.  Preliminary results of the
synthetic observations were also reported in
\citet{Tregillis01b}, \citet{Tregillis02d},
and \citet{Tregillis02c}.

Here we look anew at the simulation data presented in TJR01,
shifting our attention from a dynamical analysis to the
extraction of physical properties of the simulated objects using
standard radio and X-ray observational analyses and then
comparing results to the actual physical properties of
the simulated objects. Through this effort we aim for new insights into 
how well standard analysis assumptions and techniques
work to capture the true physical nature of a source.
In addition, we attempt to identify crucial issues for successful
extraction of physical properties of a given
object.  
We emphasize that our purpose here is not to simulate specific real
world radio galaxies.  Neither do we intend to reevaluate the
observations and analyses of particular sources.
Additional details and discussion can be found in \citet{Tregillis02e}.

The paper is structured as follows.
\S ~\ref{bic:meth} provides a brief
exposition of our numerical methods, a review of the model parameters from TJR01, 
and an overview of our synthetic observation techniques.  These techniques are then 
applied to the models in \S\S~\ref{bic:rad-x}-\ref{bic:bme}.  In
\S~\ref{bic:rad-x} we look at what can be learned from the correlations
between X-ray and
radio surface brightness.  \S~\ref{bic:bic} is devoted to an
examination of the magnetic field inferred from synchrotron and
inverse-Compton surface brightness, $B_{ic}$ and
the minimum-energy field, $B_{me}$.
Combined use of $B_{ic}$ with the  minimum-energy field $B_{me}$ as a possible
tool to estimate local nonthermal energy content is discussed 
in \S~\ref{bic:bme}.  The key findings are then summarized in \S~\ref{bic:conc}.  

\section{METHODS} \label{bic:meth}
\subsection{Numerics}  \label{bic:numerical}

Our numerical methods were detailed in
\citet{Jones99} and TJR01.  In short, we follow the
bulk flow through a 3D TVD Eulerian ideal MHD scheme and evolve a passive
relativistic electron population on the Eulerian grid through the standard
particle kinetic equation.

Our MHD code is based on an extension of the ``Total Variation Diminishing''
scheme \citep{Harten83}, as detailed in \citet{Ryu95} and \citet{Ryu95a}.
The code preserves 
$\nabla \cdot B = 0$ at each time step using a constrained transport
scheme \citep{Dai98,Ryu98}.
We include a passive ``mass fraction'' 
or ``color tracer'', $C_{j}$, to distinguish material entering the grid 
through the jet orifice ($C_{j} = 1$) from ambient plasma ($C_{j} = 0$).

Our electron transport scheme takes practical advantage of
the mismatch between bulk flow and diffusive transport scales for 
GeV electrons of relevance to
radio and X-ray emissions within radio galaxies.  In short, the lengths and
times appropriate to the dynamics are orders of magnitude larger than
those for electrons at energies relevant to
radio synchrotron radiation.  
The electron momentum (energy) distribution, $f(p)$, is then
sufficiently broad that it can be adequately represented by a 
piecewise power-law form within a few broad bins in momentum space.
At shocks, rapid diffusive acceleration
for $\lesssim 10$ GeV electrons ensures that they will emerge ``instantaneously''
from shocks with power-law momentum distributions.
Subsequent, downstream cooling
can be treated in a straightforward way.  We
therefore divide the momentum domain into a small number $N$ of
logarithmically spaced bins, and estimate particle fluxes across
momentum bin boundaries by representing $f(p) \propto p^{-q(p)}$
within bins, where $q(p)$ varies in a regular way \citep{Jones99}.

In the test-particle limit for diffusive shock acceleration
used here,
electrons emerge from shocks with a power-law spectral index
$q=3\sigma/(\sigma-1)$, where $\sigma$ is the shock compression ratio.  
The magnetic field is sufficiently weak in these simulations that
all strong shocks behave dynamically essentially as hydrodynamic shocks.
In
accord with current understandings of collisionless shocks (\eg \citet{Kang02})
we also
inject and accelerate electrons from the thermal plasma passing through shocks.
Injection physics at shocks is still poorly understood, so 
we simply assume that a small fraction, $\epsilon$,
of the thermal electron flux through a shock becomes extended via 
diffusive shock acceleration into
the aforementioned power-law 
beginning at momenta just above the postshock electron thermal values.

\subsection{Simulated Jet Properties} \label{bic:jetprop}
Here we give a short rundown of the physical parameters of the three
simulations first introduced and analyzed in TJR01. The
models are summarized in 
Table \ref{summary.t}, which is reproduced from TJR01.

The simulated MHD flows of TJR01 are 
dynamically identical.  Each jet entered the grid with a simple ``top
hat'' velocity profile with 
$M_{j} = u_{j}/c_{a} = 80$, where $c_{a}$ is the sound speed in
the uniform ambient medium.
The jets entered the grid at $x=0$ in
gas pressure balance with the ambient medium and with a
density contrast $\eta = \rho_{j}/\rho_{a} = 10^{-2}$, giving a jet-based 
Mach number $M_j=8$.  The initial jet core radius was
$r_{j}=15$ zones, while the entire $576 \times 192 \times 192$
uniformly zoned grid is $38\frac{2}{5} r_{j} \times 12\frac{2}{5} r_{j} \times 
12\frac{2}{5} r_{j}$. In units of initial jet 
radius ($r_{j} = 1$) and ambient sound speed 
($c_{a} = (\gamma P_{a}/\rho_{a})^{1/2} = 1$, with $\gamma = 5/3$) the 
simulations stopped at $\tau_{end}=5.4$ time units when the bow shock 
reached the boundary at $x=38\frac{2}{5}$. 
In physical units, the jet radius $r_{j}=2$ kpc
and the inflow velocity 
$u_j=0.05c$.  This leads to a physical time unit $\approx 10^{7}$ 
years and a computational grid length $\approx 77$ kpc.
Open boundary 
conditions were used everywhere except at the jet orifice. 
The initial axial and background magnetic field 
($B_{x} = B_{x0};~B_{y} = B_{z}=0$) (also termed ``fiducial'' below)
exerted a magnetic pressure 1\% of the ambient gas pressure ($\beta = 10^{2}$). 
In addition to the axial component, $B_{x0}$, the in-flowing jet also carried
a toroidal magnetic
field component derived from a uniform axial current and a 
return current on the jet surface; i.e., $B_\phi = 2 \times B_{x0}(r/r_{j})$ 
for $r \leq r_{j}$.  
To break cylindrical symmetry, we added a modest wobble to the 
in-flowing jet velocity; that is, it was slowly precessed on a cone of opening 
angle $5\degr$ with five periods during the run.

TJR01 presented three idealized examples of electron transport,
designed to isolate individual transport behaviors.
Those models are also summarized in Table \ref{summary.t}. 
Here, briefly, are other vital details. Electrons were transported 
explicitly over
the momentum range $p_{0} < p < p_{N}$ with $p_{0} = 10$ and $p_{N}
\approx 1.63\times10^{5}$ (with $p$ in units of $m_ec$) for all models.
Below $p_{0}$ the distribution function $f(p)$ was continued as a power-law.
At $p_{N}$ the gradient in the distribution function slope,
$dq(p_{N})/d\ln{p}$ was continued.  Eight momentum bins
($N = 8$) were used for each simulation.  Since the simulated nonthermal
cosmic-ray electrons were passive, all results
can be scaled for other choices of $p_{0}$ ($p_{N}/p_{0}$ is
fixed).  All three transport models included adiabatic cooling and
diffusive shock acceleration, although
second-order Fermi acceleration and Coulomb losses were neglected.
In each model, the jet nonthermal electron  population entered
with a momentum index $q = 4.4$, representing 
a synchrotron spectral index, $\alpha = (q-3)/2 = 0.7$, where 
$S_{\nu} \propto \nu^{-\alpha}$.

As listed in Table \ref{summary.t} we excluded local shock electron injection 
in Models {\bf 1} (hereafter called the ``\control'') and {\bf 3} (hereafter
called
the ``\cooling''), setting
$\epsilon=0$. Those models isolated evolutionary behaviors of
pre-existing electron populations in the jet. In Model {\bf 2} the 
injection parameter $\epsilon=10^{-4}$, so we
label Model {\bf 2} the ``\injection''.  Details of
these $\epsilon$ choices are given in \citet{Jones99} and TJR01.
We note that since the electron
population in the simulation was passive, the \injection~ results could be
simply rescaled for an alternate $\epsilon$.  
The in-flowing jet nonthermal electron population in this model was
made much smaller than the other two models, so that local enhancements
of the electron populations at shocks could be isolated.
The \control~ and the \injection~ are
``adiabatic'' in the sense that electrons experience
negligible synchrotron aging. That feature eliminates convex 
spectral curvature, although 
some concave curvature can result from spatial mixing of
dissimilar power-law populations.  
Model {\bf 3}, on the other hand, includes significant radiative
aging from synchrotron and inverse-Compton processes, thus its
\cooling~ label.

In order to parameterize radiative particle aging, we defined a
characteristic synchrotron cooling time, $\tau_{s0}$, for
electrons with $p = 10^4$ in the fiducial magnetic field.
For the adiabatic Models {\bf 1} and {\bf 2}, 
we set $\tau_{s0} = 1.6 \times~10^{3}$ (compared
to $\tau_{end}=5.4$) by setting $B_{x0} = 0.39 ~\mu
$G and ignored inverse-Compton losses, as well, to ensure
negligible radiative aging for electrons of interest.
For Model {\bf 3} we made $\tau_{s0} = \tau_{end} = 5.4$
by setting $B_{x0} = 5.7~\mu $G.
Inverse-Compton losses from the CMB were 
taken into account by including a term with  $B_{cmb} = 3.2 ~\mu $G
corresponding to the current epoch.  Once again, the physical unit for 
$\tau$ in these simulations is approximately 10 Myr.

\subsection{Synthetic Observation Techniques} \label{bic:synth}
Our synthetic observation technique is straightforward.
We combine vector magnetic field and nonthermal electron
distribution data from our simulations to calculate self-consistent
radio and nonthermal X-ray volume emissivities in every zone of the computational 
grid.
In order to simplify the analysis we restrict emissions 
to zones dominated by bulk plasma originating at the jet orifice,
by requiring $C_j \ge 0.99$.
A raytracing procedure integrates along lines of sight in the
optically thin limit to
project emissions from the simulated objects on the sky for an arbitrary
orientation.  We write the output data in \textsc{fits} format, and
analyze it using conventional observational packages (\textsc{miriad} and 
\textsc{karma} \citep{Gooch95}).

Since the simulations are high resolution, the synthetic maps produced are of much
higher resolution and dynamic range than typical real observations.
To see the influence of resolution on our analysis we also have
convolved the synthesized brightness distributions with circular Gaussian
beams using the \textsc{miriad} task CONVOL to several
lower resolutions. For convenience, we placed all the objects
at a fixed luminosity distance of 100 Mpc, although the choice has no
influence on our conclusions.  For this distance, the
nominal unconvolved resolution is 0.28 arcseconds, and the
projected jet length is about 110 arcseconds.  We present results
for convolved resolutions of 3.0 and 22.0 arcseconds as well, chosen to
correspond to roughly 37 and 5 telescope beams along the jet, respectively.
These choices, while arbitrary, match qualitatively what is commonly
achieved in many real source observations.
All synthetic observation images are set to 512 $\times$ 512 pixels (not
every pixel contains a nonzero brightness value, however).

As in TJR01 we confine our discussion to one representative
point in time; namely, $t = 4.0$ in
simulation units or about $4\times 10^7$ yr.
All synthetic observations in this chapter refer to the
same source orientation on the sky as that used in TJR01,
in which the jet axis is about 45 degrees from the plane of the sky.
Our conclusions are unaffected by these choices.
The bulk flow is nonrelativistic, so no Doppler corrections have
been applied.

\subsubsection{Radio Synchrotron Emissivity} \label{bic:radio}

In each spatial zone we compute a
synchrotron emissivity based on the local vector magnetic field, {\bf B},
and
nonthermal electron distribution, $f(p)$, as evolved by our transport
scheme. As given by \citet{Jones74}, the emissivity is
\begin{equation}
j_s(\nu) = j_{\alpha 0}{4\pi e^{2} \over c}f(p_{s})p_{s}^{q}\left({\nu_{B_{\bot}} \over
\nu}\right)^{\alpha}\nu_{B_{\bot}}.
\label{synch.e}
\end{equation}
The spectral index $\alpha$ is related to the local electron
momentum index $q$ via $\alpha  = (q-3)/2$, 
$\nu_{B_{\bot}}=eB\sin{\Omega}/(2\pi m_ec)$, 
where $\Omega$ projects the local field onto the sky, and
$j_{\alpha 0}$ is an order-unity dimensionless constant, defined
in \citet{Jones74}.  For a selected
observing frequency, $\nu$, the distribution,
$f(p_{s})$, and the index, $q$, are determined for each point on
the grid by establishing the
relevant electron momentum from the relation
$p_{s} = [2\nu/(3\nu_{B_{\bot}})]^{(1/2)}$, with $p_s$ in units $m_ec$.
We note for the magnetic fields in our
simulated objects and radio frequencies in the GHz band that typical 
$p_{s} \sim 10^4~-~10^5$.

\subsubsection{X-ray Emissivities} \label{bic:xray}
We compute an X-ray emissivity, $j_{X}$,
including inverse-Compton contributions from the CMB (Hereafter IC/3K)
and synchrotron self-Compton (SSC); viz,
$j_{X} = j_{3K} + j_{SSC}$.
We ignore inverse-Compton scattered AGN photons,
as well as X-ray synchrotron emission.
Although we include SSC emission, since its analysis does provide
some useful insights, we mention that it is typically several orders
of magnitude less intense in our simulated objects than IC/3K emission.

The X-ray IC/3K emissivity, $j_{3K}(\nu_X)$, can be simply
expressed at a selected X-ray frequency, $\nu_X$,
in terms of the synchrotron emissivity, $j_s(\nu_X)$, in equation (\ref{synch.e})
extrapolated to $\nu_X$; namely \citet{Jones74}, 
\begin{equation}
j_{3K}(\nu_X) = e_{\alpha 0}^{3K}~{\sigma_{T} c \over e^{2}}~
{c u_{\mu} \nu_{\mu}^{\alpha - 1} \over 4 \pi \nu_{B_{\bot}}^{1+\alpha}}~
(1 + z)^{3 + \alpha}~
j_s(\nu_X)~,
\label{cmb}
\end{equation}
where $u_{\mu} = aT^4_0$ and $\nu_{\mu}=kT_0/h$ are the energy 
density and characteristic frequency, respectively, at the current epoch of the
CMB, while $e_{\alpha 0}^{3K}$ is another order-unit constant
that can be obtained from \citet{Jones74}.
We note in equation (\ref{cmb}) that $j_s(\nu_X)$ is normalized
and $\alpha$ is determined at $p_{\mu} = (\nu_X/\nu_{\mu})^{1/2}$.
For $h\nu_X = 1.2~-~7.5$keV, considered below, $p_{\mu} \approx 2\times 10^3~-~5\times 10^3$,
which is substantially less than momenta responsible for the
GHz synchrotron emission.

The SSC emissivity, $j_{SSC}$, depends upon the synchrotron intensity
distribution incident
upon each zone.  To keep the calculation manageable, we adopt the common
approximation of an isotropic incident intensity, so that $j_{SSC}$ can be
expressed in terms of the omnidirectional incident flux, $\Phi^S_{\nu}$.
Except when the radiation field is 
dominated by a very intense anisotropic local 
source, this approximation should be
good, so adequate for our present purposes.
Then, using a convenient expression in terms of the
IC/3K X-ray emissivity from the same electron population we
have \citep{Jones74},
\begin{equation}
j_{SSC}(\nu_X) = e_{\alpha 0}^{SSC-3K}
\left(\frac{\Phi^{S}_{\nu_{k}}\nu_{k}^{\alpha_k}}{cu_\mu\nu_\mu^{\alpha-1}}\right)
\ln \left[\frac{2}{3} p_{N}^{4} \frac{\nu_{B_{\perp}}}{\nu_{X}} \right]
j_{3K}(\nu_X)~.
\label{bic:jssc.e}
\end{equation}
Here $\alpha_k$ is the synchrotron spectral index at the low-frequency
synchrotron cutoff $\nu_k$, and $\alpha$ is the 
spectral index determined for equation (\ref{cmb}).
The constant $e_{\alpha 0}^{SSC-3K}$ is order-unity and obtainable from \citet{Jones74}.
We note that since the dominant synchrotron emission occurs at much
lower frequencies than $\nu_{\mu}$, electrons responsible for
SSC emission are generally of higher energy than those producing the
IC/3K emission. 

We use a simple and fast FFT-based scheme to estimate the omnidirectional synchrotron
flux $\Phi^{S}_{\nu}$.  Details of the calculation can be found in Appendix
\ref{app:ssc}.


\section{RESULTS: RADIO AND X-RAY BRIGHTNESS CORRELATIONS} 
\label{bic:rad-x}
\begin{figure}
\begin{center}
\epsscale{0.55}
\plotone{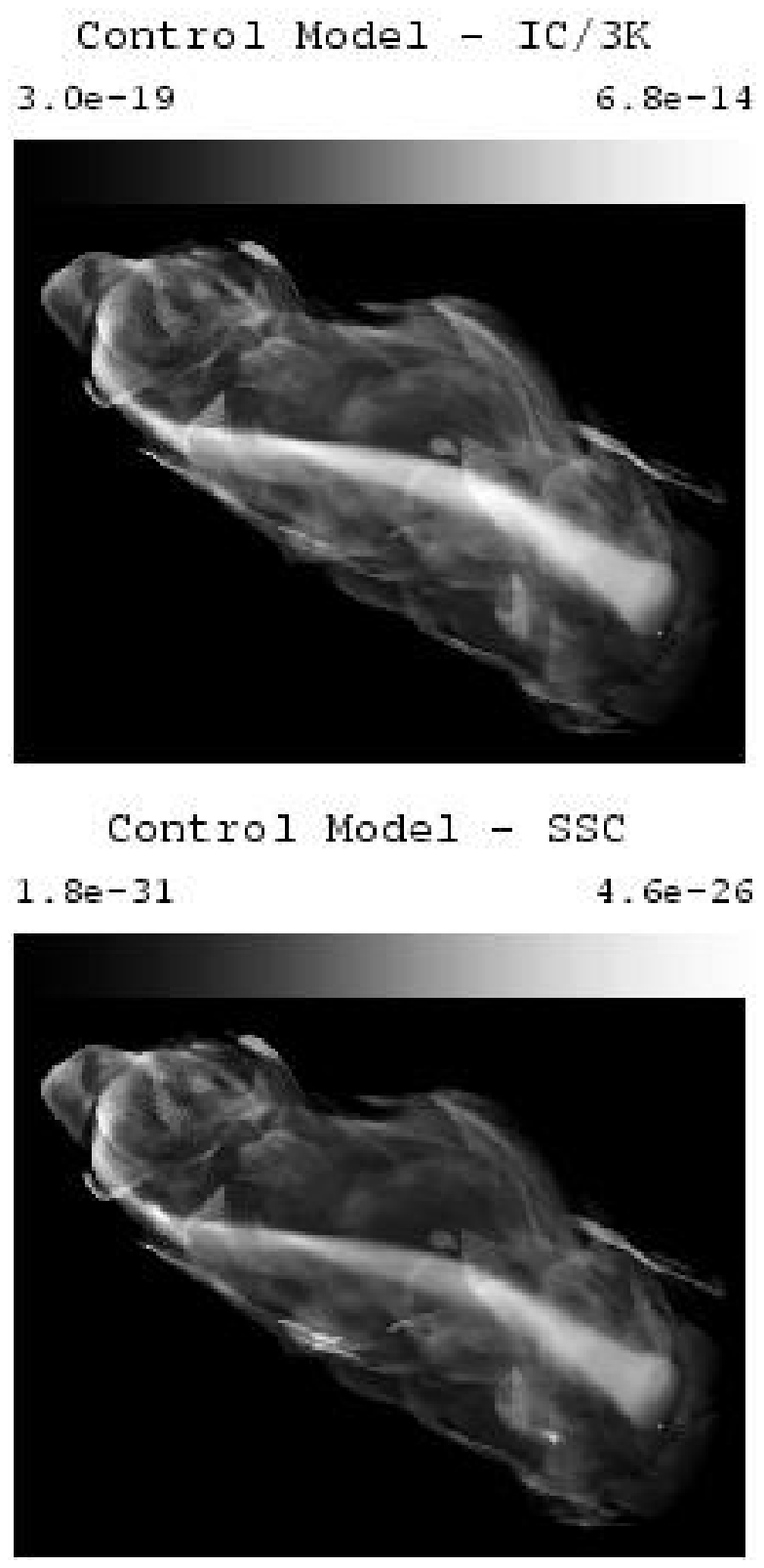}
\figcaption[f1.ps]
{1.2 keV X-ray surface brightness maps for the \control.  Both images are
displayed using a square-root transfer function.  The brightness limits 
are in code units.
\label{xray-m8.f}}
\end{center}
\end{figure}

\begin{figure}
\begin{center}
\epsscale{0.55}
\plotone{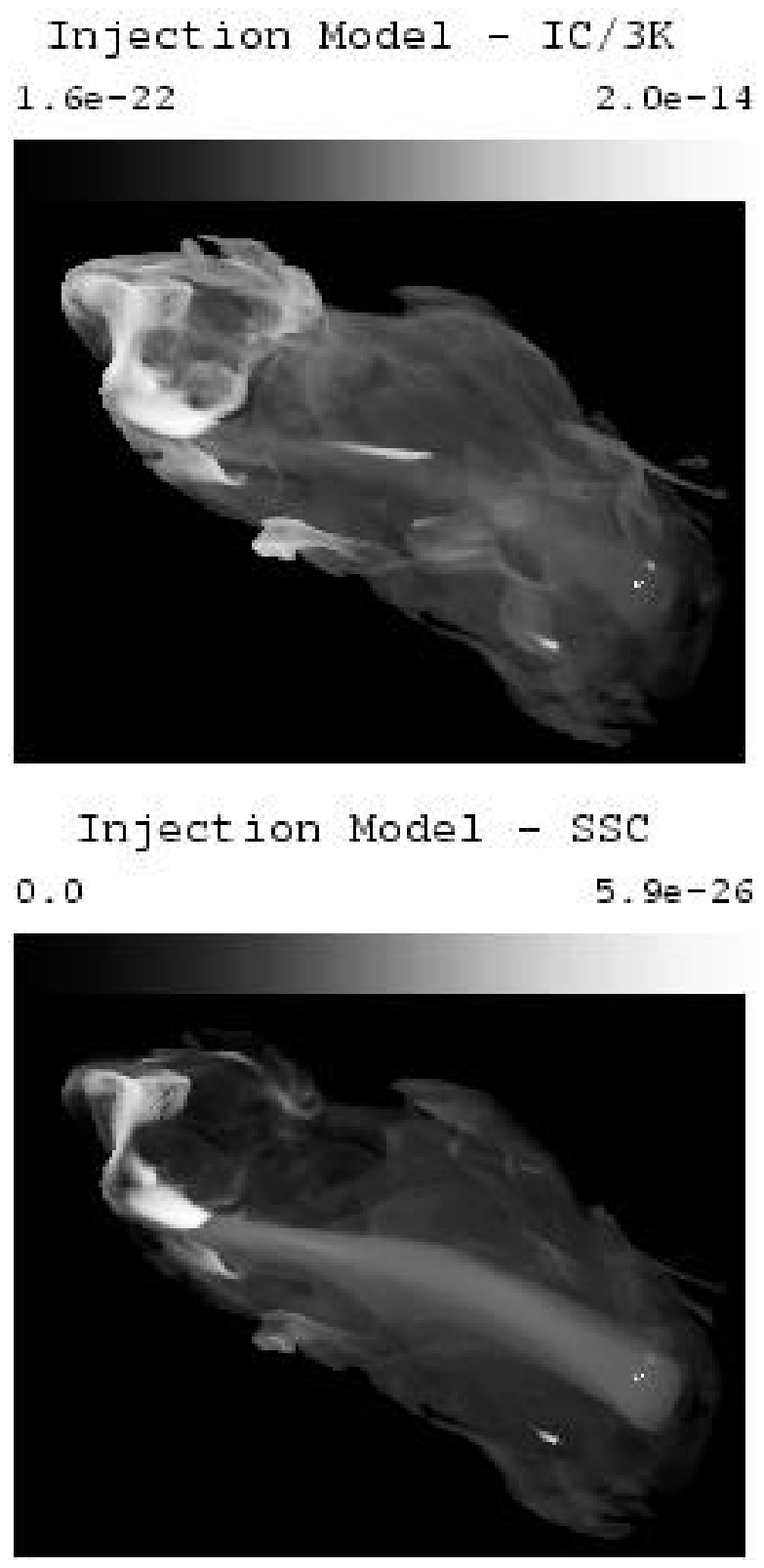}
\figcaption[f2.ps]
{1.2 keV X-ray surface brightness maps for the \injection.  Both images
are displayed using a logarithmic transfer function.  The brightness limits
are in code units.
\label{xray-m6.f}}
\end{center}
\end{figure}

\begin{figure}
\begin{center}
\epsscale{0.55}
\plotone{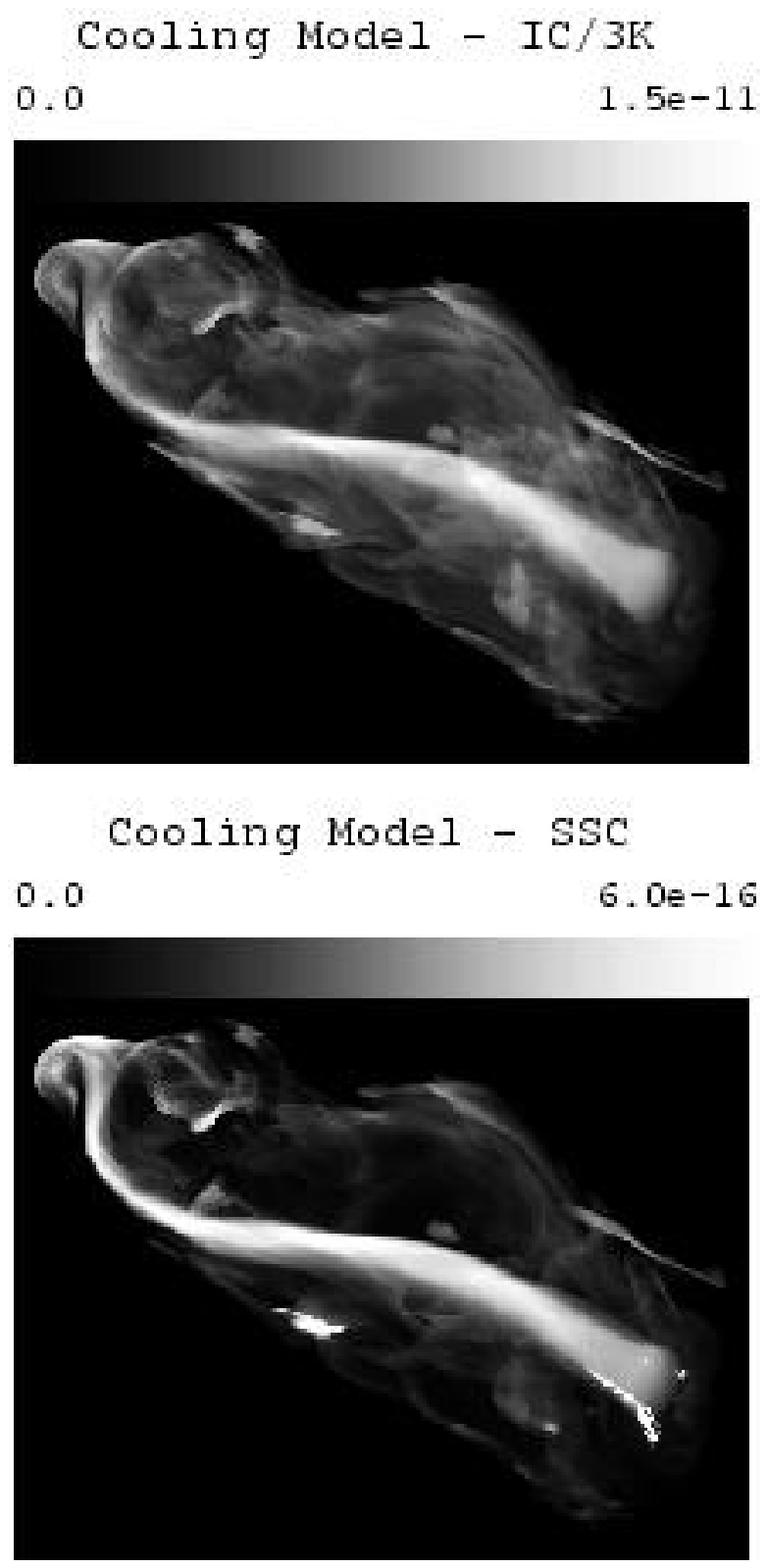}
\figcaption[f3.ps]
{1.2 keV X-ray surface brightness maps for the \cooling.  Both images are
displayed using a square-root transfer function.  The brightness limits 
are in code units.
\label{xray-m7.f}}
\end{center}
\end{figure}

We begin with a brief, qualitative comparison between the
synthetic radio and X-ray brightness distributions.
Radio surface brightness maps for the three models were shown in TJR01.
Here, Figures \ref{xray-m8.f} through
\ref{xray-m7.f} show the corresponding IC/3K and SSC X-ray
surface brightness images.
From the images it is apparent that dominant
dynamical features such as the jet and the radio hotspots are 
visible in both bands, but also that many details differ. For instance,
X-ray brightness distributions are smoother, with much less dynamic
range than the radio distributions. This simply reflects
the fact that the inverse-Compton brightness represents only the column
density of relativistic electrons over a narrow range of
electron energies. The synchrotron brightness
also depends strongly on the (intermittent) magnetic field distribution.

\begin{figure}
\begin{center}
\epsscale{0.75}
\plotone{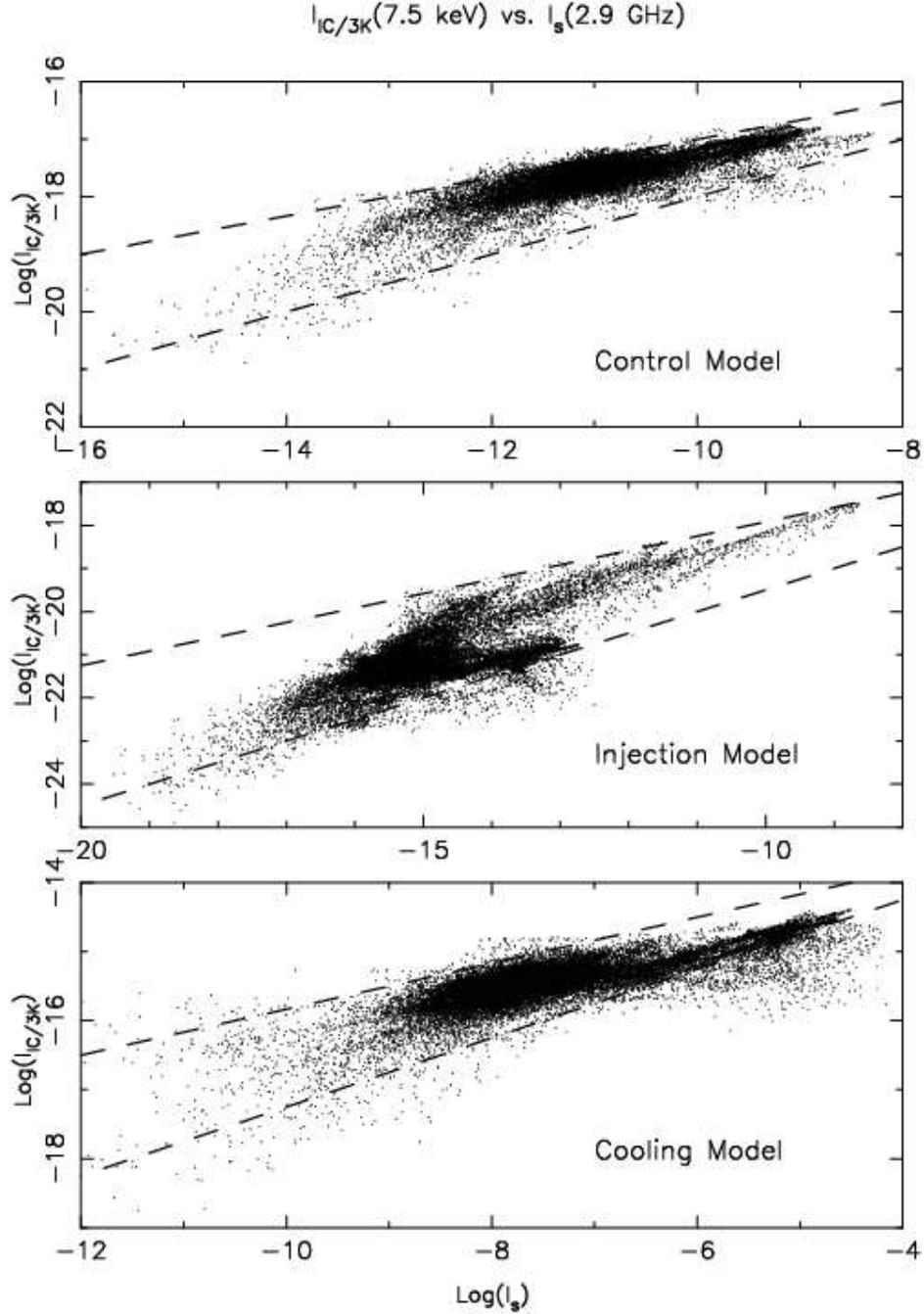}
\figcaption[f4.ps]
{Scatterplots of 7.5 keV IC/3K  vs. 2.9 GHz
synchrotron brightness in code units for each electron transport model.
In each model, the radio hotspot is represented by the diffuse collection
of points at the highest radio brightnesses.  In the \injection~
the hotspot is very prominent because it is significantly brighter than
the jet.  The brightest portions of the jet in the \control~ and \cooling~
are represented by the densest collection of points just below the
hotspot in radio brightness.  The dashed lines represent the relationships
$I_{IC/3K} \propto I_{S}^{1/3}$ and $I_{IC/3K} \propto I_{S}^{1/2}$.  They
are solely for comparison and are not statistical fits.
\label{X1vI.f}}
\end{center}
\end{figure}

\begin{figure}
\begin{center}
\epsscale{0.75}
\plotone{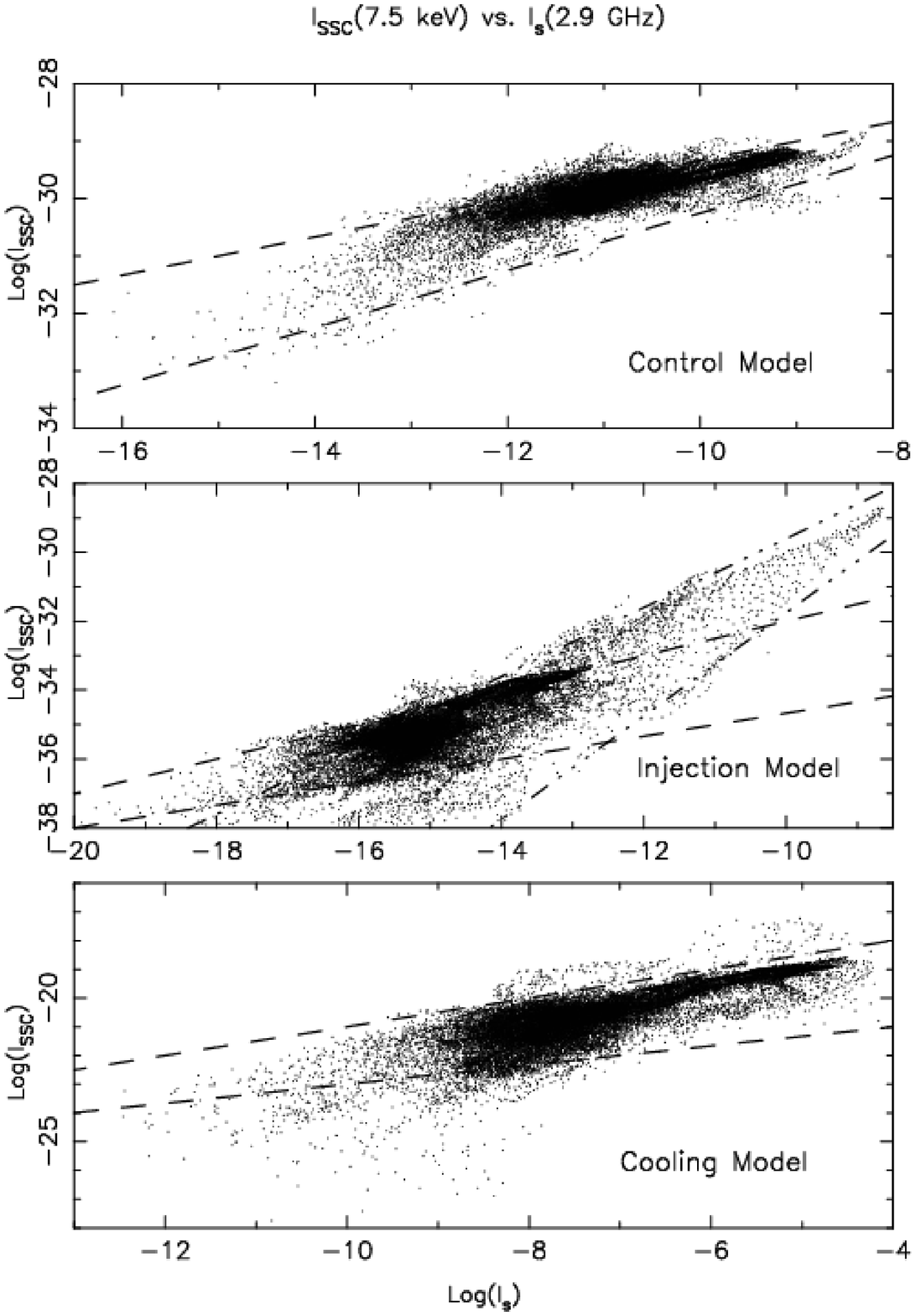}
\figcaption[f5.ps]
{Scatterplots of 7.5 keV SSC vs. 2.9 GHz
synchrotron brightness in code units. 
For numerical reasons, the minimum (log) SSC X-ray brightness is fixed at -38.
In each model the hotspot is represented by the diffuse collection of points
at the highest radio brightnesses.  The \injection~ hotspot is clearly visible.  
In the \control~ and \cooling~ the radio hotspot
is not much brighter than the surrounding jet material, represented by the dense
collection of points at slightly lower brightnesses.
The dashed lines represent the relationships
$I_{SSC} \propto I_{S}^{1/3}$ and $I_{SSC} \propto I_{S}^{1/2}$.  
The dot-dashed lines represent the relationships
$I_{SSC} \propto I_{S}^{1.0}$ and $I_{SSC} \propto I_{S}^{1.5}$. 
All lines are solely for comparison and are not statistical fits.
\label{X2vI.f}}
\end{center}
\end{figure}

\begin{figure}
\begin{center}
\epsscale{0.75}
\plotone{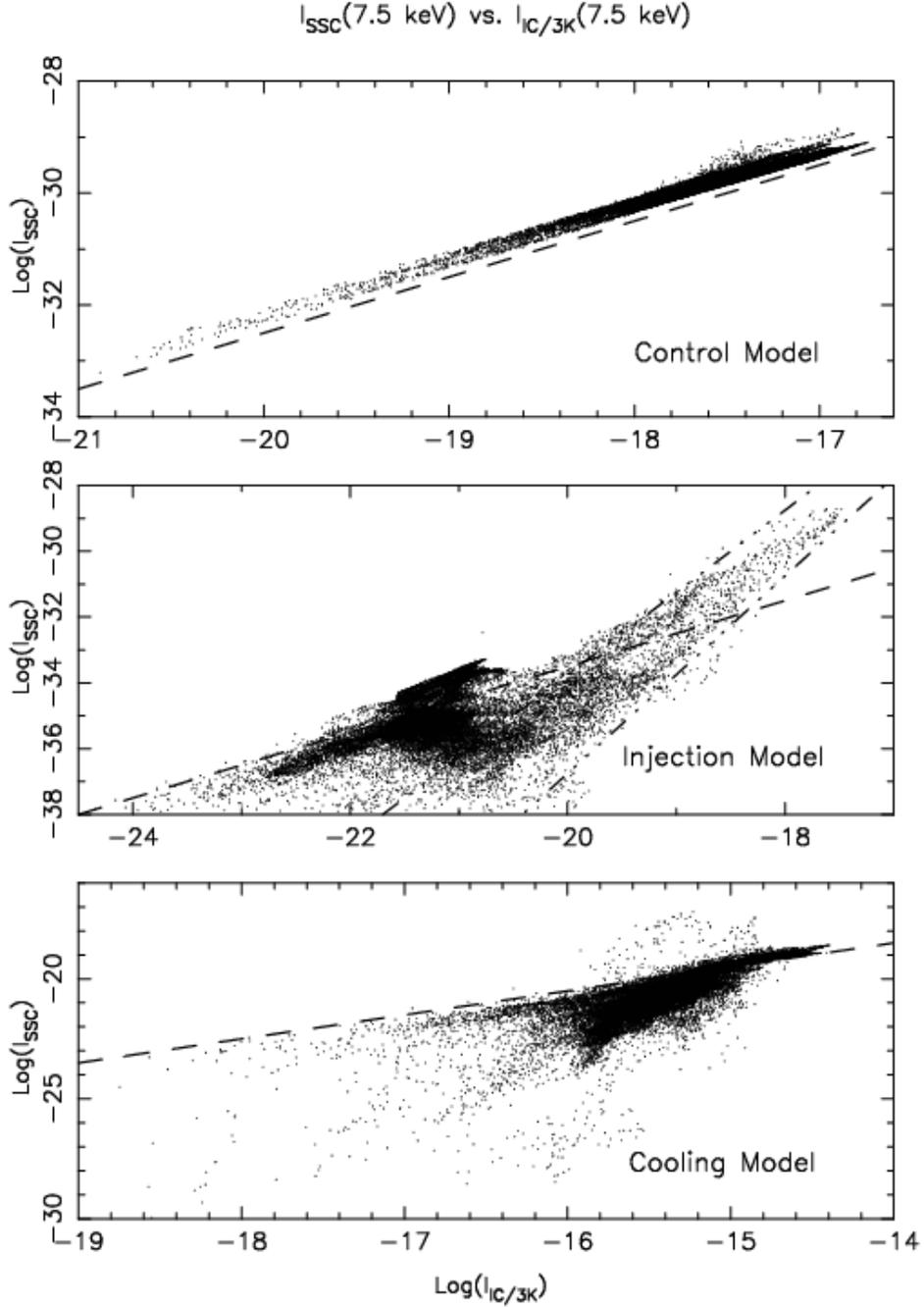}
\figcaption[f6.ps]
{Scatterplots of 7.5 keV SSC vs. 7.5 keV IC/3K brightness.
In the Control and Cooling Models, the highest SSC and IC/3K brightnesses
generally correspond to the radio hotspot, contributing to the the tight
correlation between the highest X-ray brightnesses.  The \injection~ hotspot
is again quite prominent.  
The dashed lines represent the relationship
$I_{SSC} \propto I_{IC/3K}^{1.0}$.  
The dot-dashed lines represent the relationships
$I_{SSC} \propto I_{IC/3K}^{2.5}$ and $I_{SSC} \propto I_{IC/3K}^{3.0}$. 
All lines are solely for comparison and are not statistical fits.
\label{X2vX1.f}}
\end{center}
\end{figure}

In \S\S \ref{bic:bic} and \ref{bic:bme} we will 
combine the synthetic radio and X-ray data to explore their effectiveness
for inferring source magnetic field properties. First, however,
we demonstrate that some basic dynamical relationships connecting
particles and fields can be extracted from correlations between
the various intensity distributions, which are shown in
Figures \ref{X1vI.f}, \ref{X2vI.f} and \ref{X2vX1.f}.
As we examine our three simulated objects, we should keep in mind that
the magnetic field structures are identical in all three models, except 
for magnitude; only the nonthermal electron distributions differ.
Also, we emphasize, once again,
that our goal is to examine ways of extracting reliable information
about the objects we have in hand; that is, how we extract
from observations the physical properties that are actually present.
Because we know their properties exactly, 
the simulated objects are uniquely valuable testing grounds, independent
of how closely those detailed properties match astrophysical objects.

Figures \ref{X1vI.f} and \ref{X2vI.f} are scatter plots of 
7.5 keV IC/3K and SSC intensities versus 2.9 GHz radio intensity,
respectively, for the three electron
transport models (Table \ref{summary.t}). 
Similar distributions would be obtained for other observing frequency choices.
Every other nonzero pixel is represented, for just under $2.5 \times 10^4$
points in each plot.  
In each panel the X-ray and radio brightness distributions are broadly 
correlated.
It is possible to extract from these trends insights about the dynamical behaviors
of the source magnetic fields, as we illustrate. 
We emphasize, however, that at a given radio (X-ray) 
brightness, the X-ray
(radio) intensities generally spread over more than an order of
magnitude, so it would be unrealistic to try to predict the inverse-Compton
brightness distribution from the radio distribution or vice versa. 
This is especially
so in the lobe structure, where the spread in synchrotron brightness is
enormous, due to the wide range of magnetic field strengths.

Comparison between Figures \ref{X2vI.f} and \ref{X1vI.f} shows 
that the SSC brightness distributions of our
simulated objects are generally several orders of magnitude less
than the IC/3K intensities. The difference is more than 10 orders
of magnitude for both the \control~ and \injection, where the 
magnetic field magnitude was set very small to reduce synchrotron
energy losses. Even for the \cooling, with its stronger field, the SSC
intensities are mostly three to four orders of magnitude less
than the IC/3K intensities. This behavior comes from the fact
that the synchrotron
omnidirectional flux is generally much smaller than the CMB omnidirectional 
flux in our simulated objects. That mirrors the situation in
real radio galaxies, as well, since only the most intense X-ray hotspots
have been found to be SSC dominated \citep[e.g.,][]{Harris00}.
Despite this, it is instructive to include the SSC brightness
distributions in our analysis, since they 
reveal some useful insights.

The IC/3K vs radio trends in Figure \ref{X1vI.f} are well
described for all three models by the 
proportionality $I_{IC/3K} \propto I_{S}^{m}$, with $m \approx 1/3
- 1/2$.  
To understand this, suppose the local
magnetic field scales with the electron density as $B
\propto n_{e}^{b}$.  Since $I_{IC/3K} \propto n_{e} \cal{D}$ and $I_{S}
\propto n_{e}{\cal{D}}~B^{1+\alpha}$, where $\cal{D}$ is the path length,
we have $I_{IC/3K} \propto
I_{S}^{1/(1+b(1+\alpha))} \equiv I_{S}^{m}$. 

On the other hand, by standard 
arguments $B \propto \rho^{2/3}$ for a disordered field in 
compression-dominated flows. Assuming $n_e \propto \rho$, this gives 
$m = 3/(5+2\alpha)$.
For $\alpha \sim 0.5~-~1$, typical of the synthetic radio sources,
this would give $m\sim 0.5~-~0.43$.
Alternatively, at perpendicular shocks one expects $B \propto \rho$, resulting in
$m = 1/(2+\alpha)$, or $m \sim 0.4~-~0.33$. Thus, the
observed trends are consistent with field evolution
dominated by these dynamical processes. 
On the other hand, field evolution controlled by stretching of 
flux tubes would satisfy $B \propto \ell$, where $\ell$ is
the length of the flux tube \citep[\eg][]{Gregori00}. There is
no explicit interdependence between density and magnetic field.
We know that the magnetic fields in the simulated object are
filamented, so flux tube stretching certainly takes place. 
The above exercise brings out the fact that the global field evolution
in the simulated object is dominated by compression, however.
This matches our conclusions about the
simulated field behaviors in TJR01. 
Sheared field evolution, while clearly involved and locally
important, predominantly adds 
scatter to the correlations in Figure \ref{X1vI.f}.

With the notable exception of the dominant hotspot in the \injection,
the SSC and radio intensity distributions
follow  similar correlations as the IC/3K-radio correlation;
namely, $I_{SSC} \propto I_{S}^m$, with $m \approx 1/3~-~1/2$.
From this match between SSC and IC/3K trends one might expect that 
$I_{SSC} \propto I_{IC/3K}$, and this is nicely confirmed for
the \control~ in Figure \ref{X2vX1.f}. 
This proportionality is also evident in the other two models,
excepting the \injection~ hotspot, although there is a lot more scatter.

For the general distribution we note from equation 
\ref{bic:jssc.e} that one can write $I_{SSC} \propto \Phi^{S}~I_{IC/3K}$.
The observed correlations then imply, especially for the \control, that
$\Phi^{S} \approx~constant$ within the radio lobes. That represents
the fact that the
average omnidirectional flux for the source is a good approximation
to the local flux at any point in the source. This follows if
the omnidirectional synchrotron flux at each point is dominated
by the collective contributions of distant emissions within the source.
In that case the average synchrotron emissivity becomes an effective
value to use in estimating the SSC emission from the object.

The situation is much different in the dominant \injection~ hotspot.
The SSC/radio intensity correlation
is described by $m \sim 1~-1.5$, whereas
the SSC and IC/3K intensities follow $I_{SSC} \propto I_{IC/3K}^{2.5-3}$.
On the other hand, the correlation between IC/3K and radio
surface brightness is similar to what is seen elsewhere in the source
and in the other two models.
Previously, we interpreted the IC/3K vs radio correlation as evidence 
for magnetic fields
evolving according to the scaling $B\propto n_e^{2/3}$, which, therefore, 
seems applicable in the hotspot, as well.
It is the SSC behavior that is different in the \injection~ hotspot, and the
distinctive correlations there can be understood by supposing that
$\Phi^S \propto I_S$; that is, the synchrotron radiation field
is dominated by local emissions in the hotspot, as we might expect. 
Using again the emissivity relations in \S 2.3 we can write in
this case that $I_{SSC} \propto I_S^{(8+2\alpha)/(5+2\alpha)}$
and $I_{SSC} \propto I_{IC/3K}^{(8+2\alpha)/3}$. Taking $\alpha \approx 0.65$
for the \injection~ hotspot radio spectral index we would
predict $I_{SSC} \propto I_S^{1.47}$ and $I_{SSC} \propto I_{IC/3K}^{3.1}$,
very close to what is observed.

Finally, we note that on the whole SSC X-ray emission is no more tightly
correlated with the radio intensity than is the IC/3K intensity,
despite the fact that it is a consequence of the radio emission itself.
That comes from the fact that the SSC emissivity reflects radio emissions 
throughout the source, not just the local radio emission.
Small regions of high radio brightness embedded in more diffuse
emission tend to wash out the SSC/radio correlation, because they act
like internal point sources that anisotropically illuminate their
neighborhoods.  If the surrounding emission is particularly diffuse, such
regions may contribute flux at considerable distances throughout the
source.  That effect is particularly striking in 
the \injection, where the very bright radio hotspot is compact and
surrounded by more diffuse radio emission.

\section{RESULTS: ESTIMATED MAGNETIC FIELD STRENGTHS FROM OBSERVED INTENSITIES}
\label{bic:bic}

As mentioned in the introduction, total synchrotron intensities are 
commonly applied in two ways
to estimate magnetic field strengths in
optically thin synchrotron sources. This is not
straightforward, of course, since synchrotron intensity from
an optically thin source alone cannot determine the strength of
the source magnetic field. Some additional information must be
supplied. 
The more traditional and still most-common method minimizes
the total energy (or sometimes pressure) for the relativistic particles 
and magnetic field, or what is close to the same thing, an 
equipartition is assumed between them.
A second, and increasingly common method combines information
carried in the synchrotron intensity with observed 
inverse-Compton scattered intensities presumably produced
by the same electron population. 
There are serious uncertainties in the application of both methods, ranging from
the absence of convincing theoretical arguments for the minimum
energy condition in an unrelaxed system to the likelihood of 
inhomogeneous magnetic field and particle
distributions. In this section we will apply standard methods for
these two approaches to synthetic observations of our simulated
objects in order to see how they compare and to understand better
how the inferred field properties match actual conditions
for objects with complex, quasi-realistic structures.

\subsection{The Analysis Procedure}

\begin{figure}
\begin{center}
\plotone{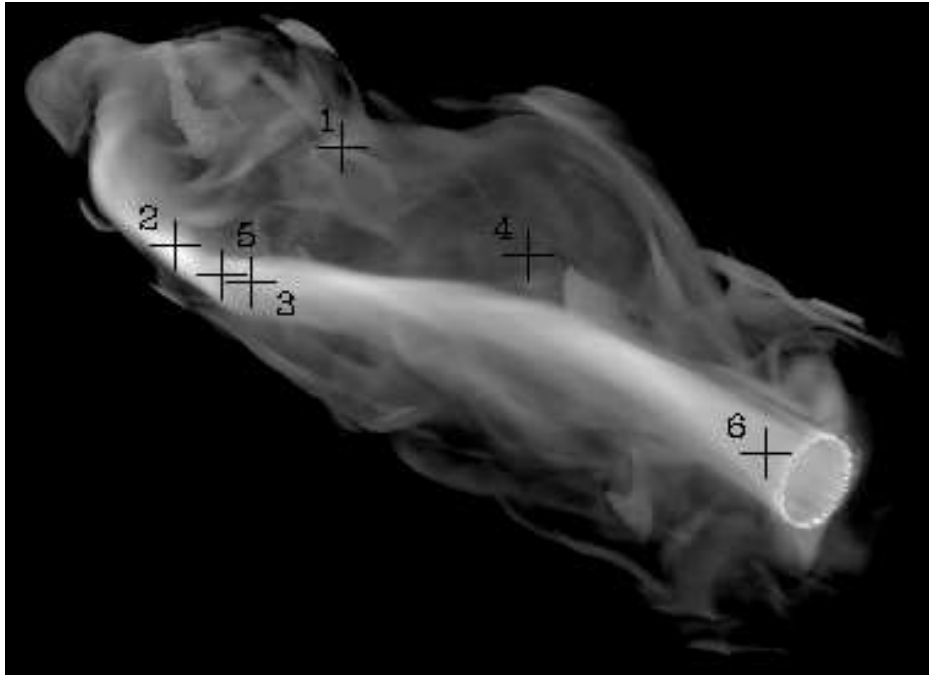}
\figcaption[f7.eps]
{Greyscale image of the \control~ 8 GHz synchrotron surface brightness.
The 6 crosses denote the lines of sight (LOS) used for 
analysis of the IC/3K and energy partitioning in \S\S \ref{bic:bic}-\ref{bic:bme}.
\label{regions.f}}
\end{center}
\end{figure}

\begin{figure}
\begin{center}
\epsscale{0.80}
\plotone{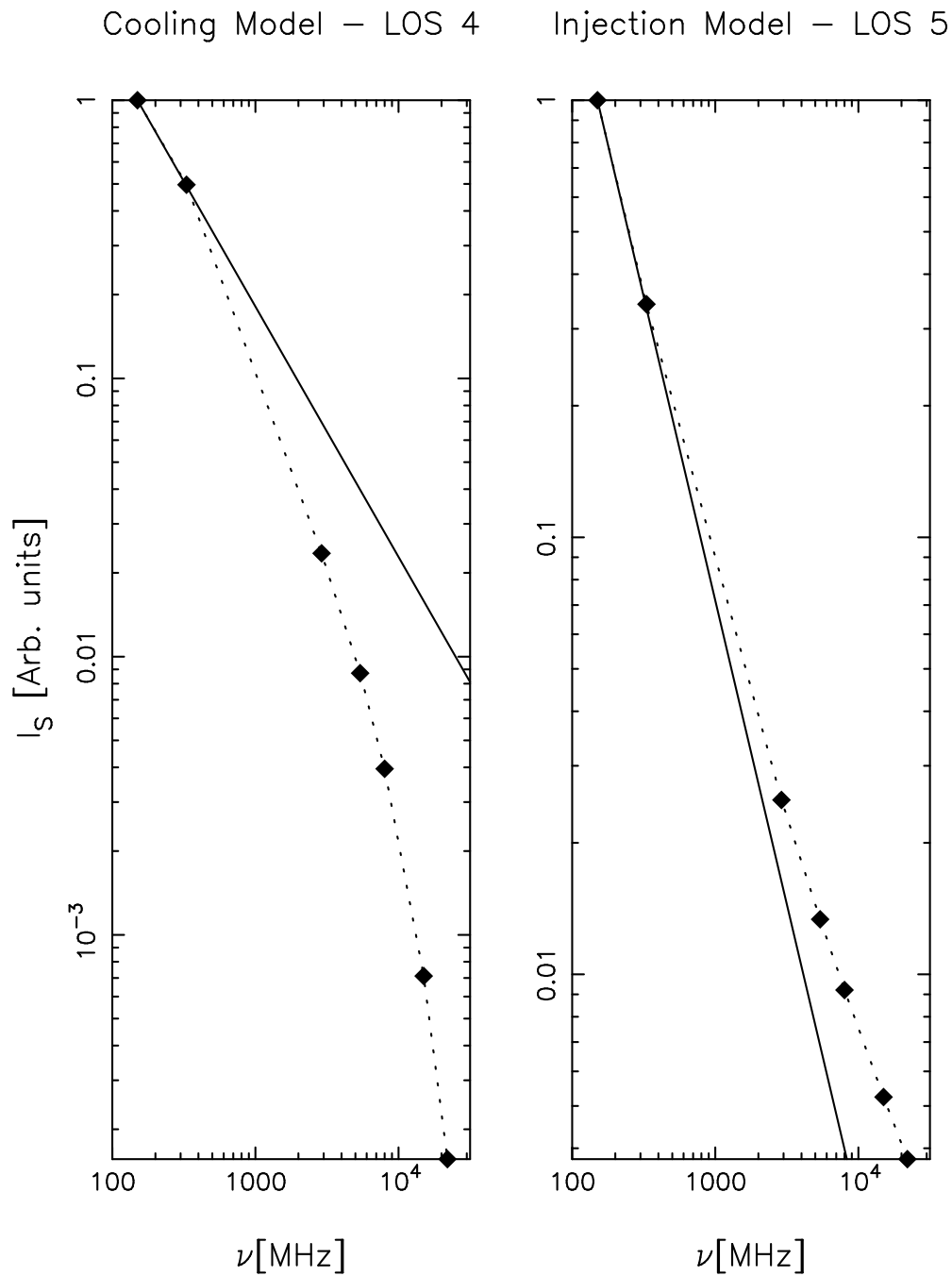}
\figcaption[f8.eps]
{Synchrotron spectra at {\bf LOS 4} in the Cooling Model and {\bf LOS 5} in the
Injection Model.  Both spectra are comprised of synthetic surface brightness
calculations at 150 MHz, 330 MHz, 2.9 GHz, 5.4 GHz, 8.0 GHz, 15.0 GHz, and
22.0 GHz.  The spectra have been scaled so that the surface brightness is equal
to unity at the lowest frequency.  For reference, we have included a power-law 
line with $\alpha = 0.90$ 
in the left panel, and a power-law line with $\alpha = 1.39$ 
in the right panel.  Note that in both cases, significant deviations from the
power-law form are apparent at frequencies as low as 2.9 GHz.
\label{losspec.f}}
\end{center}
\end{figure}

For this exercise we ``observed'' synchrotron emission at $\nu = 2.9$ 
GHz and IC/3K emission at 1.2 keV in all three models.
To explore the connections between observationally
inferred magnetic fields and actual magnetic field
distributions we selected six lines of sight (LOS)
for close analysis. Those LOS are marked in Figure \ref{regions.f}.
Two of the LOS pass through lobe structures ({\bf LOS 1}, {\bf LOS 4}). 
Three penetrate the jet ({\bf LOS 3}, {\bf LOS 5}, {\bf LOS 6})
and one centers on the dominant hotspot ({\bf LOS 2}). 
{\bf LOS 4} allows us to look at the influence of particle
aging on the analyses, since the spectrum there is strongly convex in
the \cooling. Similarly, {\bf LOS 5} shows a concave spectrum
in the \injection, due to strong blending of dissimilar power-law
electron distributions introduced at shocks.  
(These spectra along {\bf LOS 4}
and {\bf LOS 5} are shown in Figure \ref{losspec.f}.)
{\bf LOS 6} passes through 
the jet base, where magnetic field and particle populations are particularly
simple, and known from the jet inflow conditions.

Magnetic field strength can be calculated very simply from synchrotron
and IC/3K intensities for a uniform 
medium when the electron distribution is a power-law. We used the following
expression for this estimated field, $B_{ic}$
\citep[\eg][]{Jones74,Harris74,Harris79}:
\begin{eqnarray}
&&\nonumber B^{1+\alpha}_{ic} =
(1.06 \times 10^{-11})~(2.09\times10^{4})^{\alpha-1}(1+z)^{3+\alpha}\\
&&~~~~~~~~\times\left(\frac{\nu_r}{\nu_X}\right)^{\alpha}~
\left(\frac{j_{\alpha 0}^{3K}}{j_{\alpha 0}}\right)~
\frac{I_S(\nu_r)}{I_{IC/3K}(\nu_X)}~~ \mu{\rm Gauss}.
\label{bic.e}
\end{eqnarray}
For uniform particle and field distributions, 
this expression is exactly equivalent to inverting the analytic calculation of 
synchrotron and IC/3K surface brightnesses in our synthetic observations.

Similarly, the standard expression used to compute the minimum-energy magnetic field, $B_{me}$, 
is (\citet{Miley80})
\begin{eqnarray}
&&\nonumber B_{me} = 5.69 \times 10^{-5} 
\left[ ~\left(~\frac{1+k}{\eta}\right)~\frac{F(\nu_{r})}{\nu_{r}^{-\alpha}} 
\frac{(1+z)^{3+\alpha}}{\theta_x~\theta_y~\ell~
\sin^{3/2}{\vartheta}}\right]^{2/7}\\
&&~~~~~~~\times\left[
\frac{\nu_{2}^{1/2 - \alpha} - \nu_{1}^{1/2 - \alpha}}
{(1/2 - \alpha)}
\right]^{2/7}~{\rm Gauss}.
\label{bme}
\end{eqnarray}

Here $F(\nu_r)$ is the observed radio flux density within an observing beam,
$k \equiv U_{proton}/U_{E}$,  
$\eta$ is a magnetic field volume filling factor, $\vartheta$ is
the angle between the magnetic field and the line of sight, $\theta_x$
and $\theta_y$ are the semimajor and semiminor axes of the observing
beam in arcseconds, $\ell$ is the path length through the source in kpc,
and $\nu_{1}$ and $\nu_{2}$ are the {\it fixed} lower and upper synchrotron 
cutoff frequencies in the source frame, expressed in GHz. Note that
while $B_{me}$ is expressed in terms of synchrotron flux density, it
really depends on the mean $I_S(\nu_r)/\ell$; that is, the mean synchrotron emissivity.
We assume below for simplicity that $k = \eta = \sin{\vartheta} = 1$.
The magnetic field is so tangled and intermittent that assuming
$\sin{\vartheta} = 1$ introduces errors of only a few percent 
into our analysis.  

In order to incorporate the full nonthermal
electron distributions as well as possible into
$B_{me}$, the frequency limits needed in
equation (\ref{bme}) correspond in each model to the 
characteristic
synchrotron frequencies of the lowest and highest energy electrons
and the fiducial magnetic field for each model (see Table \ref{summary.t}).
Those turn out to be $\nu_1 = 100$ Hz and $\nu_2 = 30$ GHz for the
\control~ and the \injection; for the \cooling~ $\nu_1 = 1500$ Hz and $\nu_2 = 450$ GHz.
Our conclusions do not depend on any of these parameter choices.

We note that since both empirical field measures depend only
on intensities, they are independent of distance assumptions.
Nevertheless, for
consistency with our discussions in TJR01 we place the objects
at a distance of 100 Mpc, giving 0.28 arcseconds for
the projected size of a numerical resolution element.

\begin{figure}
\begin{center}
\epsscale{0.90}
\plotone{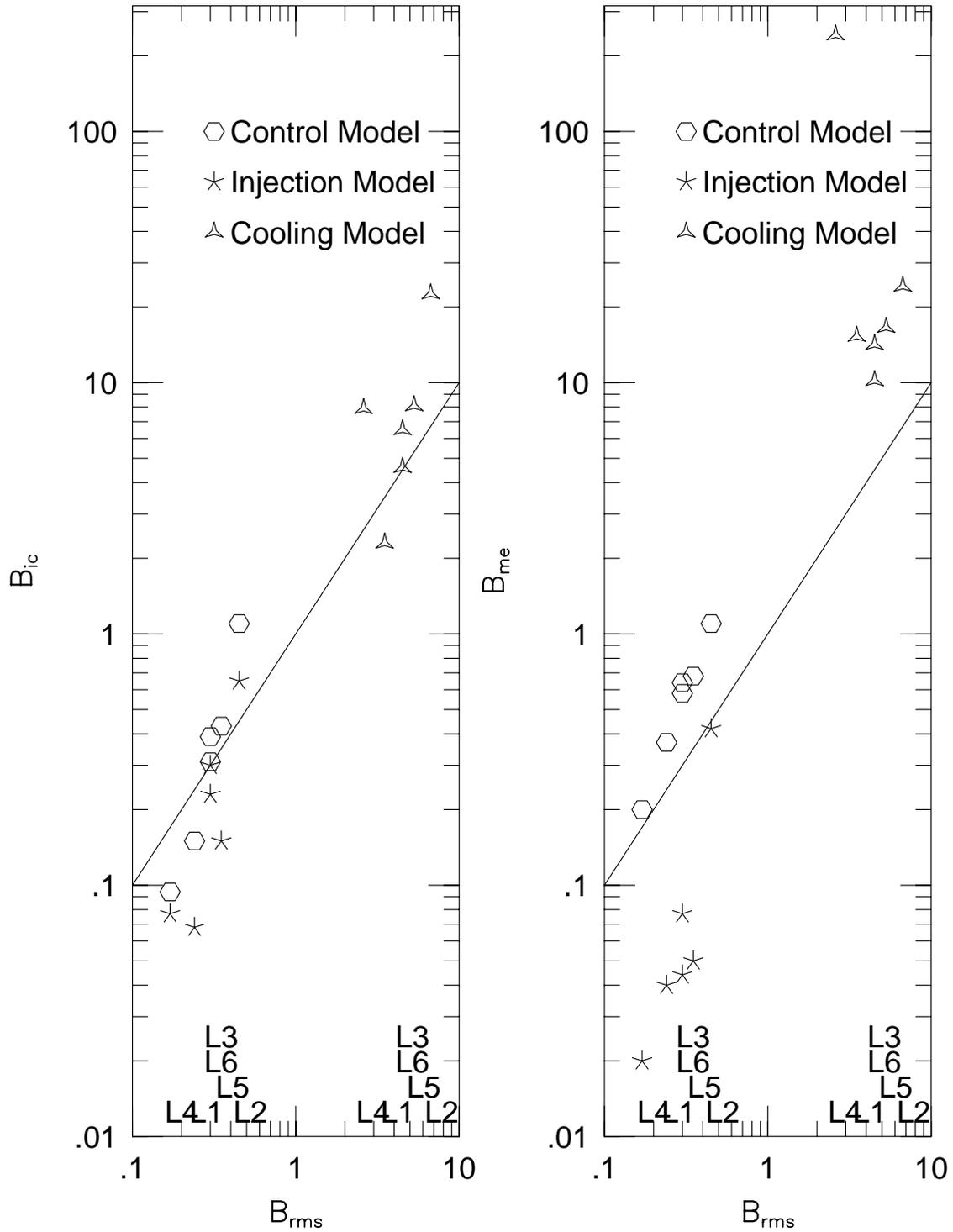}
\figcaption[f9.eps]
{Comparisons of magnetic field measures from Table \ref{bic1.t}. Symbols
indicate the electron transport model. Associated LOS numbers
are marked at the bottom of each plot according to the associated $B_{rms}$ 
values.
\label{bcomb.f}}
\end{center}
\end{figure}

Summaries of our analysis
are given in Table \ref{bic1.t} and Figure \ref{bcomb.f}.  At high resolution,
$B_{ic}$ generally falls within a factor of two of $B_{rms}$ in the sampled
volume.  $B_{me}$, on the other hand, can show
considerably larger scatter, over an order of magnitude in some cases, depending
on the specifics of the particle transport.  Both estimates are sensitive to
spectral curvature.  Estimates for jet structures tend to show less scatter
than for lobe structures, which typically embody a wider range of physical
conditions.  Results for 3.0 and 22.0 arcsecond Gaussian beams are also 
included in Table \ref{bic1.t}.
The influences of bigger beams are predictable
in that derived fields resemble the analogous properties of the
larger regions surveyed. 

Figure \ref{bcomb.f} compares $B_{ic}$ and $B_{me}$ to $B_{rms}$.
It is evident that both empirical magnetic field estimates correlate
roughly with $B_{rms}$ along the selected lines of sight. 
The bolometric synchrotron intensity does depend on the rms
magnetic field along the line of sight, of course. 
The spectral emissivity $j_s(\nu) \propto B^{1+\alpha}$,
is similar, since in our sources $\alpha \sim 0.7~-~1$ are common.
Still it is not obvious a priori how well such a simple
measure as $B_{rms}$ should compare to inferred values, so the
experiment is valuable.
Two-thirds
of the $B_{ic}$ ($B_{me}$) points are within a factor of two (three)
of $B_{rms}$. On the other hand, while the $B_{ic}$ values are 
approximately randomly distributed with respect to $B_{rms}$,
there are obvious biases in $B_{me}$ that are dependent on the 
electron transport model in the simulation. Those biases correctly reflect
actual deviations from the minimum energy condition. In particular,
our simulated objects are not in electron/magnetic field
energy equipartition. 
Nor, as we will discuss in \S \ref{bic:bme}, is there physics in the
simulations expected to produce this kind of equipartition.

Magnetic field
profiles for each of the LOS are shown in Figure \ref{blos1.f}
for the \control~ and \injection. The field distributions in
the \cooling~ are identical to the other two sources
except for being everywhere a factor 14.8 times larger.
However, since the particle populations evolve differently
in the \cooling, the inferred magnetic field behaviors
are not necessarily the same as in either of the other two models.
Points where $C_j < 0.99$ are set to zero in this analysis; that is, we
filtered out plasma that did not originate in the jet in order to
match the assumptions made in computing nonthermal emissions.
In this regard we comment that the contact discontinuity between 
the ``jet plasma''
and the ``ambient plasma''; \ie the boundary between $C_j = 1$ and
$C_j = 0$, is not simple. It is actually quite convoluted, so that 
distinct fingers of ambient plasma penetrate into
the cocoon. Consequently, there are ``dropouts'' in the emission
and associated physical variables according to the prescription we have
followed; that is, these regions do not
contribute to $B_{ic}$, $B_{me}$, nor to $B_{peak}$
or $B_{rms}$.

In \S \ref{bic:bme} we will 
explore how effectively $B_{ic}$ and $B_{me}$ can be used to reveal
information about the partitioning of energy between magnetic fields
and nonthermal electrons. Here we focus on the two field estimators 
separately.
We summarize the important features in the next subsection.

\subsection{Field Properties on Selected Lines of Sight \label{bic:los.ss}}
\begin{figure}
\begin{center}
\epsscale{0.75}
\plotone{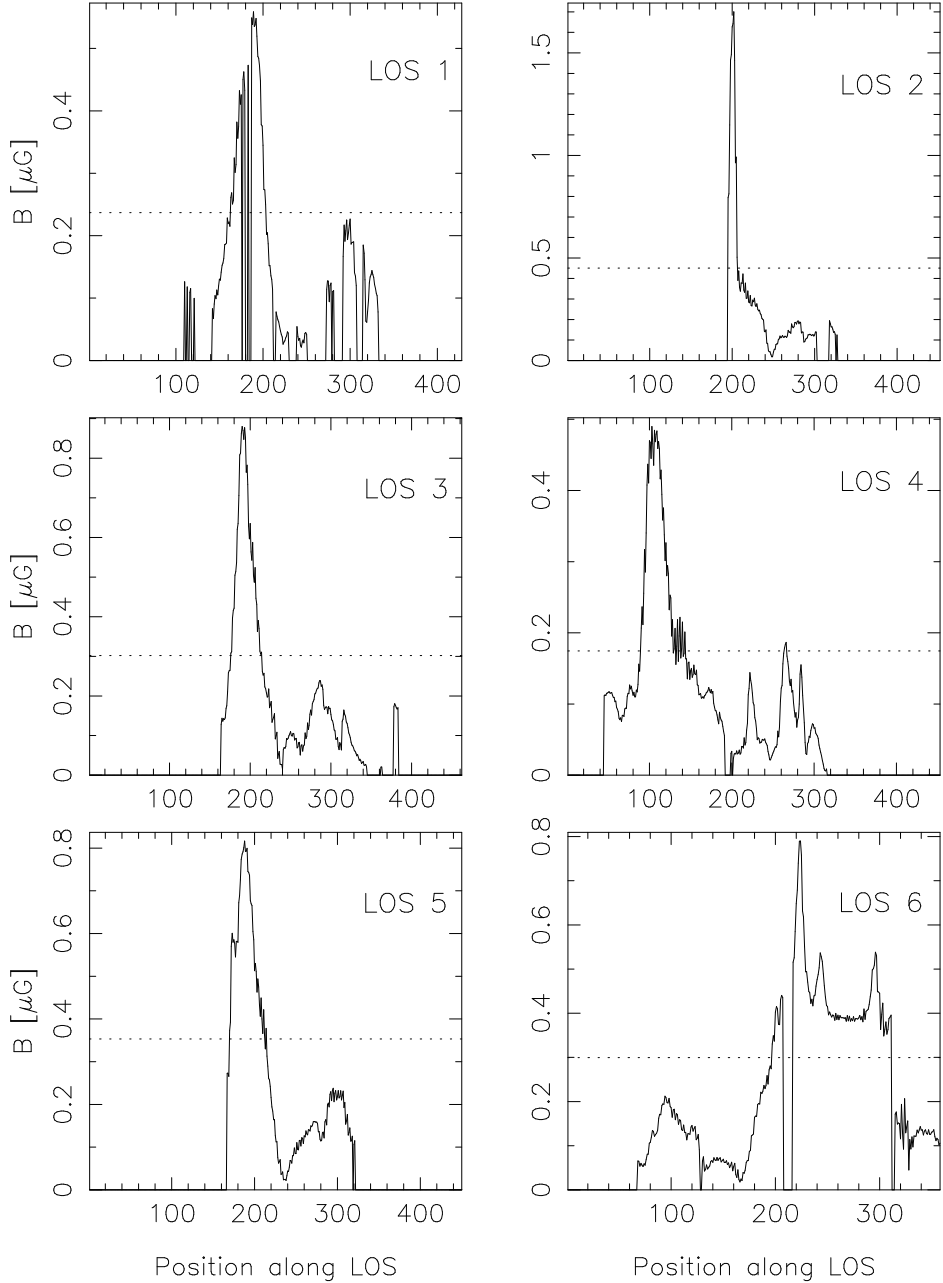}
\figcaption[f10.eps]
{Physical magnetic field values along {\bf LOS 1-6} marked in Figure
\ref{regions.f} for the \control~ and \injection.  For the
\cooling~ increase each field point by a factor 14.6. Wherever
the jet mass fraction (``color'') is less than 99\% the fields
have been filtered out.
On each LOS the filtered RMS field
is marked with a dotted line; the RMS values are
listed in table \ref{bic1.t}. The distance unit is 
computational zones along the LOS.  Low values correspond to the
side of the source closer to the hypothetical observer.
\label{blos1.f}}
\end{center}
\end{figure}

In this section we look directly at the properties of the magnetic field
along the individual selected LOS as revealed in Figure \ref{blos1.f}.

{\bf LOS 1} and {\bf LOS 4} pass through the diffuse
lobe of the sources. Both lines exhibit very complex magnetic fields,
and considerable entrainment of ambient plasma
into the lobe structure is evident from the large number of
``dropouts'' in Figure \ref{blos1.f}. The strongest field
regions are revealed as the relatively bright 
filaments in Figures \ref{xray-m8.f}-\ref{xray-m7.f} and
Figure \ref{regions.f} (see also Fig. 5 in TJR01). The nonthermal
electron density also exhibits much fine structure (not shown),
as one would expect from the discussion in \S \ref{bic:rad-x}.
While the IC/3K-radio analysis (\S \ref{bic:rad-x})
indicates the field and particle structures are correlated,
it also shows very wide scatter, especially in the lobes, where flows are chaotic. That
does impact on $B_{ic}$ as a quantitative
field measure.

Except for one case ({\bf LOS 4} in the \cooling) values of $B_{ic}$ 
along these lobe LOS all fall below $B_{rms}$. 
This bias results from the presence of substantial
electron populations in weak field regions that contribute little to the
radio emission, but that do produce X-rays. 
In effect the IC/3K intensity provides
an overestimate of the number of radio emitting electrons, so
that under the uniform source hypothesis
the field apparently needed to account for the radio emissions is weakened.
On the other hand, a comparison with Figure \ref{blos1.f}
shows that $B_{ic}$ for the \control~ and \injection~ are 
representative of the field values being sampled along both LOS,
so it does give a ``meaningful'' result, if not a simply
defined quantitative one.

The relatively higher value of $B_{ic}$ for the \cooling~ on {\bf LOS 4}
is a consequence of spectral aging; that is, the
spectrum is convex. Intentionally, no correction was made
for this effect, in order to expose its potential influence.
Recall that
the 1.2 keV IC/3K emission used to compute $B_{ic}$ comes from 
1 GeV electrons, while the 2.9 GHz emission is
produced by electrons with energies about an order of
magnitude higher, even in the strongest field regions along
this LOS. With a convex electron spectrum the
computed $B_{ic}$ will be artificially increased, as simple
arguments can show. Suppose, for example, we measured the
bolometric synchrotron intensity, $I_S$, from electrons of energy $\gamma_S$
and the bolometric IC/3K intensity, $I_{3K}$, from electrons of energy $\gamma_{3K}$.
If the object were homogeneous, and we fixed the observed radio band, 
it is simple to show that $B \propto (I_S/I_{3K}) N_{\gamma_{3K}}/N_{\gamma_S}$,
where $N_{\gamma_{3K}}$ and $N_{\gamma_S}$ represent the number of
electrons required to produce the observed intensities. If we assumed
that $N_{\gamma_{3K}}$ and $N_{\gamma_S}$ were connected by a powerlaw
in $\gamma$ with the powerlaw index determined by the high energy electrons
responsible for the radio band,
but the distribution is actually convex, we would always
overestimate $B$, since we would overestimate $N_{\gamma_{3K}}/N_{\gamma_S}$.
This same influence has an even more striking impact on the
value of $B_{me}$ for the \cooling~ on {\bf LOS 4}. Under the
equipartition that accompanies the minimum energy assumption, the 
effective overestimate of the electron population also exaggerates the 
estimated magnetic field energy, leading to an estimated magnetic
field about two orders of magnitude greater than $B_{rms}$
and an order of magnitude greater than $B_{peak}$ along this LOS.
Thus, it is very important when analyzing sources that exhibit
``aged'' spectra to account for the curvature when calculating
the electron energy (I. L. Tregillis and L. Biggs, in preparation).

The values of $B_{me}$ found in the \injection~ along these LOS, by
contrast are almost an order of magnitude below the RMS field values.
This correctly reflects the fact that the magnetic field energy in this
model actually does greatly exceed the nonthermal electron energy in most
locations that generate synchrotron emission.

{\bf LOS 3}, {\bf LOS 5} and {\bf LOS 6} intersect the jet. 
$B_{ic}$ values are close to $B_{rms}$ in each model when the
highest resolution is used in the observations. For the \control~and \cooling~
the inferred field is almost independent of
angular resolution. The \injection~ jet is
relatively less dominant in accounting for emissions, 
since the radiating electron population is small there.
At high resolution the inferred field is also close to
$B_{rms}$. The concave property of the synchrotron
spectrum on {\bf LOS 5} in the \injection~ has at most a modest
depressing influence on $B_{ic}$. 
As beam size increases on {\bf LOS 3} and {\bf LOS 5}, $B_{ic}$ in
the \injection~ increases markedly, because emissions become
influenced by the nearby hotspot, where the electron population
increases dramatically. That influence is 
greater in the \injection~ since its hotspot is much
more intense than in the other two models (see Fig. \ref{X1vI.f}).

$B_{me}$ values follow a similar pattern to those seen in $B_{ic}$ for lobe LOS;
that is, the relation between $B_{me}$ and $B_{rms}$ follows
the actual relationship between nonthermal particle and
magnetic field energies in the predominant emission regions.

{\bf LOS 6} passes through the jet near its origin. In Figure \ref{blos1.f}
the jet is contained roughly between position coordinates
245 - 295, measured in computation zones. The magnetic
field structure is simple and approximates that introduced at
the computational boundary. Indeed, for all three models $B_{ic}$
is very close to $B_{rms}$ as well as the projected axial
base jet magnetic field in the simulation. 

{\bf LOS 2} passes through a hotspot resulting from
a shock produced as the jet impinges on the near boundary of its cocoon.
The magnetic field is relatively compressed and organized there,
accounting for the dominant peak in the field distribution near
line coordinate 200. Virtually
all of the radio emission in this direction originates in the
hotspot, so that the $B_{ic}$ and $B_{me}$ values for all three models lie 
close to $B_{rms}$, irrespective of angular resolution. This is the only
LOS we sampled that returns a $B_{me}$ estimate in the \injection~ that
is close to $B_{rms}$. It is, in fact, the only LOS we sampled that
has emissions in that electron transport model predominantly
from regions close to electron/magnetic field equipartition.

\section{RESULTS: OBSERVATIONAL ESTIMATION OF PARTICLE/FIELD ENERGY PARTITIONING}
\label{bic:bme}

In the previous section we found that a synchrotron/inverse-Compton
analysis provides reasonable estimates of characteristic magnetic
fields, in particular $B_{rms}$, in our simulated objects.
There are biases that can degrade the estimates when the fields
are significantly more intermittent than the nonthermal
electrons or when the electron energy spectra are strongly
aged. However, the $B_{ic}$ values we derived were mostly
within a factor of two of the RMS fields. On the other hand
the minimum energy magnetic fields were close to the RMS fields (or
any other simple statistical measure) only when the emission
was dominated by relativistic plasma close to equipartition. That is
what we should hope for and expect, of course. At the same time, we noted
that the biases in $B_{me}$ correctly reflected the degree of
deviation from equipartition of the dominant plasma. 

This suggests that combining $B_{ic}$ and $B_{me}$ might reveal
a meaningful estimate of the ratio of nonthermal electron and
magnetic field energy and, consequently, provide an improved
estimate of the total nonthermal energy compared to the 
minimum energy. We test that possibility in this section for
our simulated objects, where we not only can perform
radio and X-ray observations, but also know the internal
energetics directly.

\subsection{The Analysis Procedure}

The obvious parameter for comparing magnetic and nonthermal particle
energies is the ratio of their local densities; namely,
\begin{equation}
d \equiv \frac{U_{B}}{U_{E}}.  
\label{db.e}
\end{equation}

Standard expressions for the minimum-energy magnetic field, $B_{me}$,
(\ie assuming a uniform source and a power-law $f(p)$,
and a fixed frequency band) (see eq. [\ref{bme}]),
lead to the simple relation 
\begin{equation}
d = \frac{3}{4} (1+k) ~\left( \frac{B}{B_{me}} \right)^{7/2}~,
\label{dbme.e}
\end{equation}
where $B$ is the actual field strength.
The exact value of $k$ makes no substantial difference to our
conclusions, so for convenience we will still
apply the commonly used value $k = 1$. 
The total energy density in nonthermal particles and magnetic fields,
$U_T$, is then obviously
\begin{eqnarray}
\nonumber
&&U_T = U_B (1 + \frac{1 + k}{d})\\
&&= \frac{3}{7} U_{me} \left[\left(\frac{B}{B_{me}}\right)^2
+\frac{4}{3}\left(\frac{B_{me}}{B}\right)^2\right]~,
\label{utot.e}
\end{eqnarray}
where $U_{me} = (7/3)(B_{me}^2/(8\pi))$ is the combined minimum energy
density.

It is convenient to define $d_{min} \equiv \frac{3}{4}(1+k)$ corresponding to
$B = B_{me}$. For $k = 1$ this gives $d_{min} = 1.5$.
Using $d_{min}$ we can write simply
\begin{equation}
d = d_{min}\left(\frac{B}{B_{me}}\right)^{7/2}~,
\label{dbme2.e}
\end{equation}
or
\begin{equation}
\frac{B}{B_{me}} = \left( \frac{d}{d_{min}} \right)^{2/7}.
\label{dbme3.e}
\end{equation}
Note for the discussion below, when $d \gg 1$ ($d \ll 1$), that $U_T \propto d^{4/7} U_{me}$ 
($U_T \propto d^{-4/7} U_{me}$).

To set the stage for our discussion we note that
the jet inflow boundary conditions discussed in \S \ref{bic:jetprop} 
give $d_0 = 0.16$ for
the \control~ and the \cooling. 
Thus, these two jets enter the grid relatively
close to electron/field equipartition. 
Accordingly, from equation (\ref{dbme3.e}) one finds at the
jet orifice that $B \approx 0.5 B_{me}$.
In contrast, the \injection~
jet enters very far from equipartition, with $d_0 = 1.6 \times 10^3$, 
so $B \approx 7.3 B_{me}$.

How should we expect $d$ to change from these input values
as the flows evolve? Recall that in our simulations 
particle energies are dominated by the thermal population; the
nonthermal electrons are ``test particles'' that respond to the
underlying flow dynamics, which is based on otherwise fully consistent
MHD.
The expected variation of $d$ under adiabatic 
compression in those flows is straightforward to estimate.
Recall from \S \ref{bic:rad-x} that compression of
an isotropic magnetic field varies $B \propto \rho^{2/3}$, so that 
$U_{B} \propto \rho^{4/3}$.  We found in \S \ref{bic:rad-x}
that this behavior seems to account for much of the
field structure in the lobe. At the same time, the nonthermal electron gas is
relativistic, so $U_E \propto \rho^{4/3}$, as well. 
Thus, we may expect adiabatic expansion or compression to produce relatively
little change in $d$. 

On the other hand, in sheared flows
$B$ may have little or no correlation with $\rho$.
In particular, an incompressible stretched flux tube has a
field strength depending only on the length of the tube (see \S \ref{bic:rad-x}),
which could produce very large increases in $d$ inside flux tubes. 

Variations of $d$
across shocks are also difficult to quantify, although qualitatively we
can expect $d$ to decrease there, perhaps by a large factor. 
At most $U_B \propto \rho^2$ (in a perpendicular shock), 
but $U_E$ should increase more strongly than this in response
to diffusive shock acceleration.
As shock strength increases the jump in $U_B$ increases asymptotically,
but the jump in $U_E$ does not, both due to the diffusive reacceleration of
the incident particle flux and also to any fresh injection due to ``thermal
leakage'' at the shock.

\subsection{Comparison Between Inferred and Actual Energetics}\label{bic:bme-anal.ss}
Because of initial conditions,
most regions within the \control~ and \cooling ~{\it coincidentally} lie within an
order of magnitude of this equipartition condition,
with nonthermal electron energy being typically
somewhat greater than magnetic field energy 
(see Figs.~\ref{dlos-m8.f},\ref{dlos-m7.f}). 
In contrast for much of
the source volume in the \injection, magnetic field energy greatly exceeds
the nonthermal electron energy density (Fig. \ref{dlos-m6.f}). 
That contrasting condition simply reflects the small population of
nonthermal electrons entering with the jet in this model. It 
conveniently provides an excellent opportunity to study how
different energy balances between particles and
fields reveal themselves through their emission properties.

\begin{figure}
\begin{center}
\epsscale{0.90}
\plotone{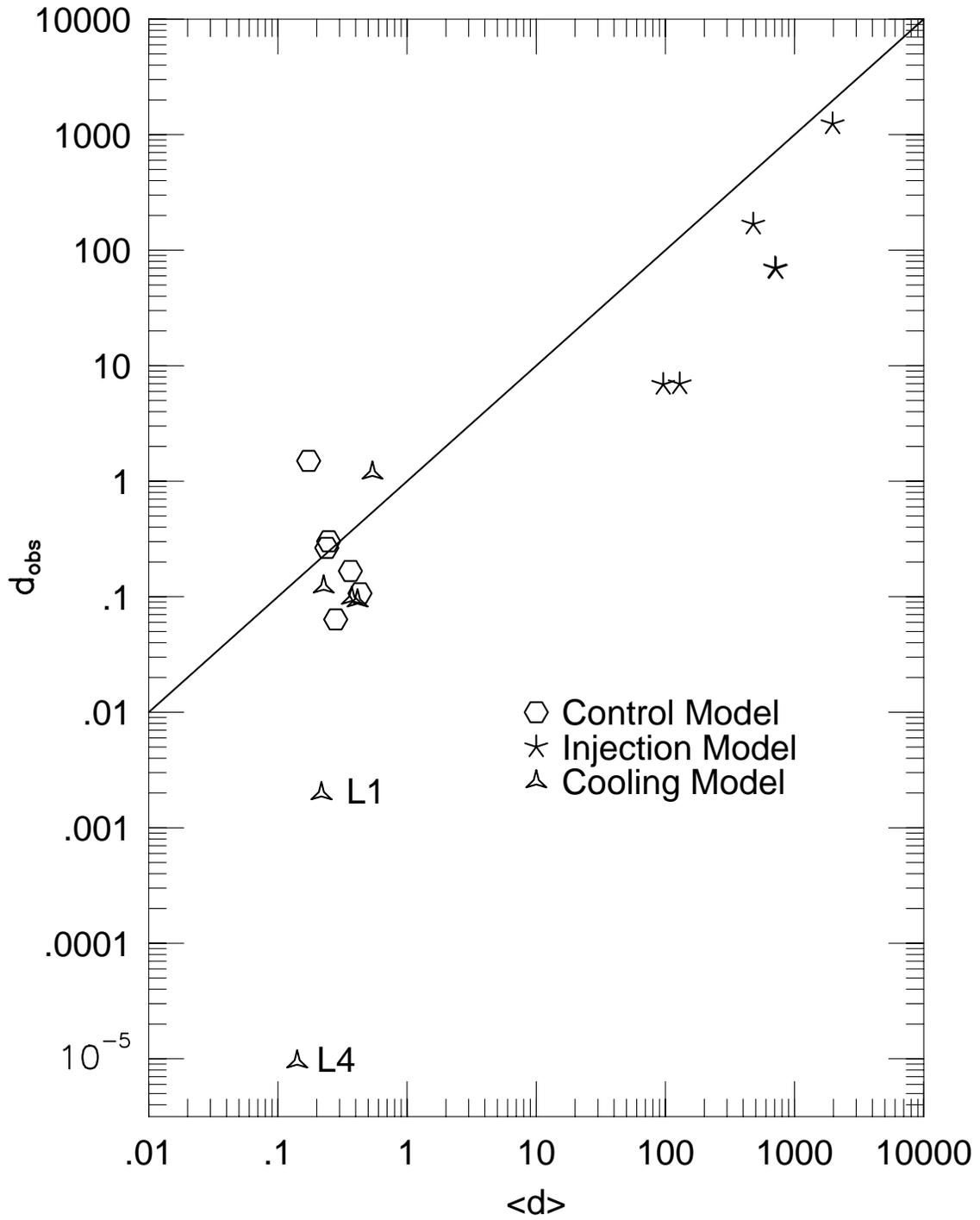}
\figcaption[f11.eps]
{Comparisons of $d_{obs}$ and $\left< d \right>$ from Table \ref{dtable.t}. Symbols
indicate the associated electron transport model.
\label{epart.f}}
\end{center}
\end{figure}

\begin{figure}
\begin{center}
\epsscale{0.75}
\plotone{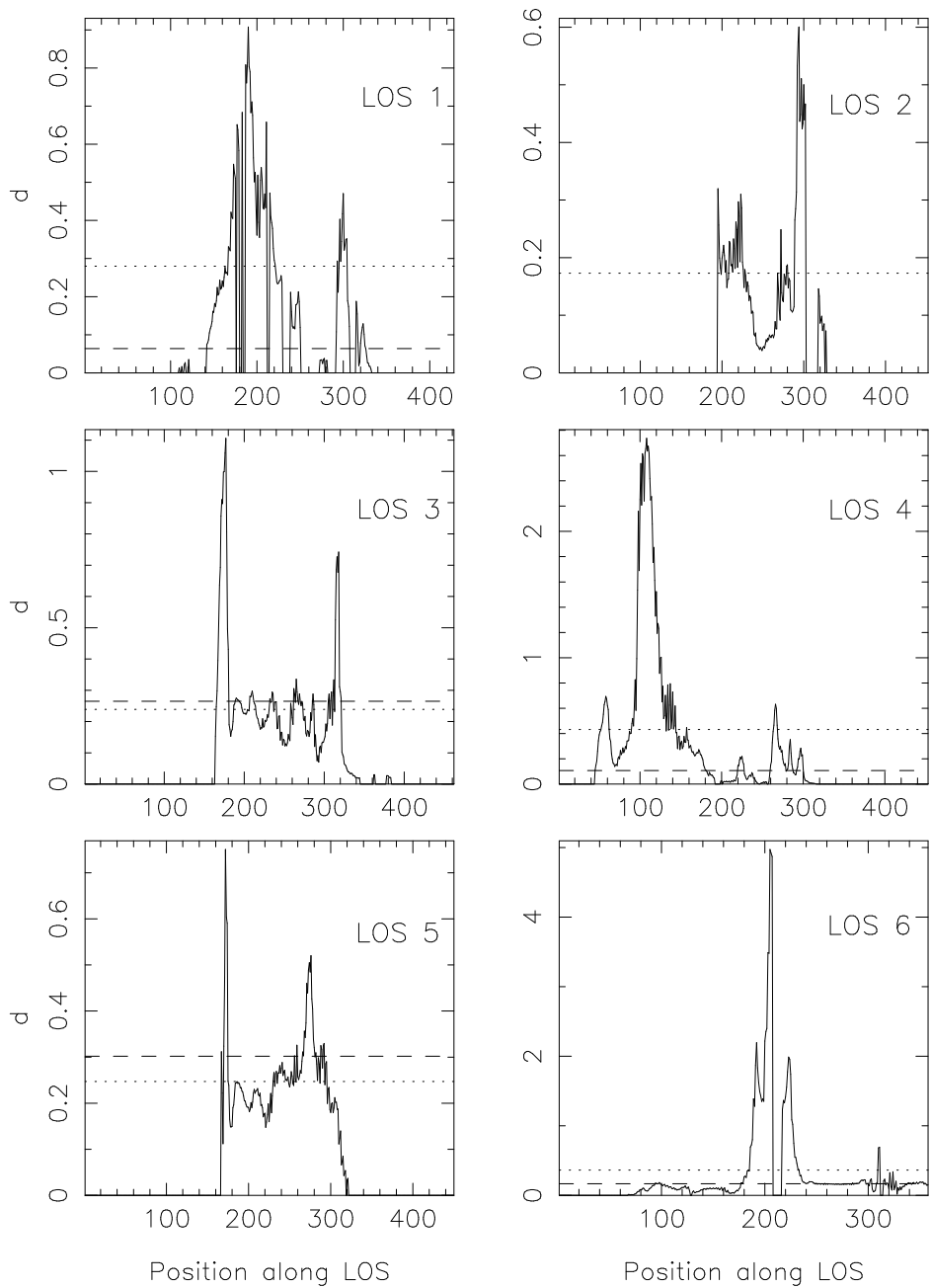}
\figcaption[f12.eps]
{Parameter $d$ along {\bf LOS 1-6} for the \control.
The average $\left< d \right>$ 
is marked by a dotted line, and the ``observed'' value from equation
(\ref{dbme2.e}) with $B=B_{ic}$ is marked with a dashed line.  
For reference, on axis the in-flowing jet carries $d_0 = 0.16$.
Distance refers to
computational zones along the LOS.  Low values correspond to the
side of the source closer to the hypothetical observer.
\label{dlos-m8.f}}
\end{center}
\end{figure}

\begin{figure}
\begin{center}
\epsscale{0.75}
\plotone{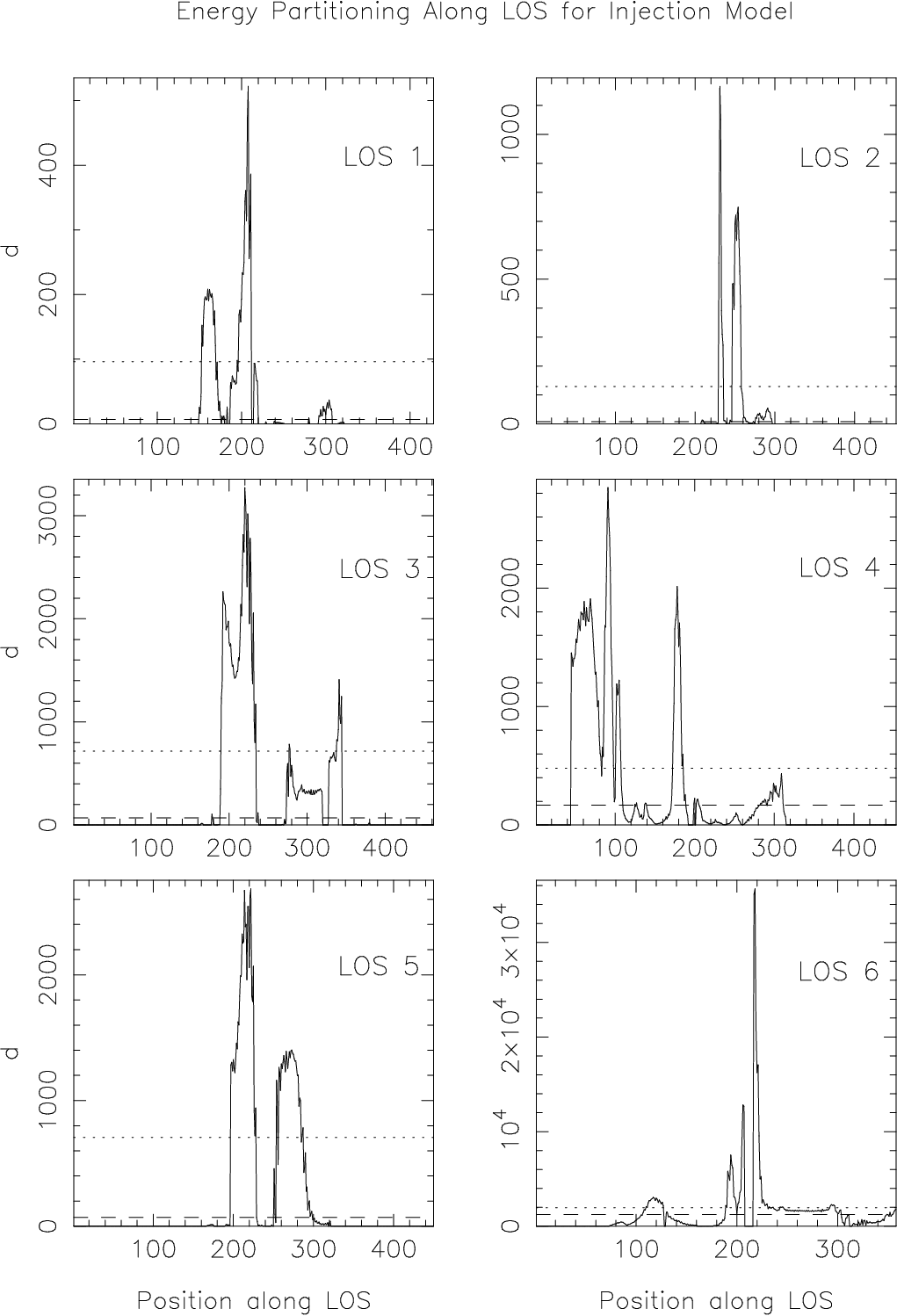}
\figcaption[f13.eps]
{Same as Figure \ref{dlos-m8.f} for the \injection.
For reference, on axis the in-flowing jet carries $d_0 = 1600$. 
\label{dlos-m6.f}}
\end{center}
\end{figure}

\begin{figure}
\begin{center}
\epsscale{0.75}
\plotone{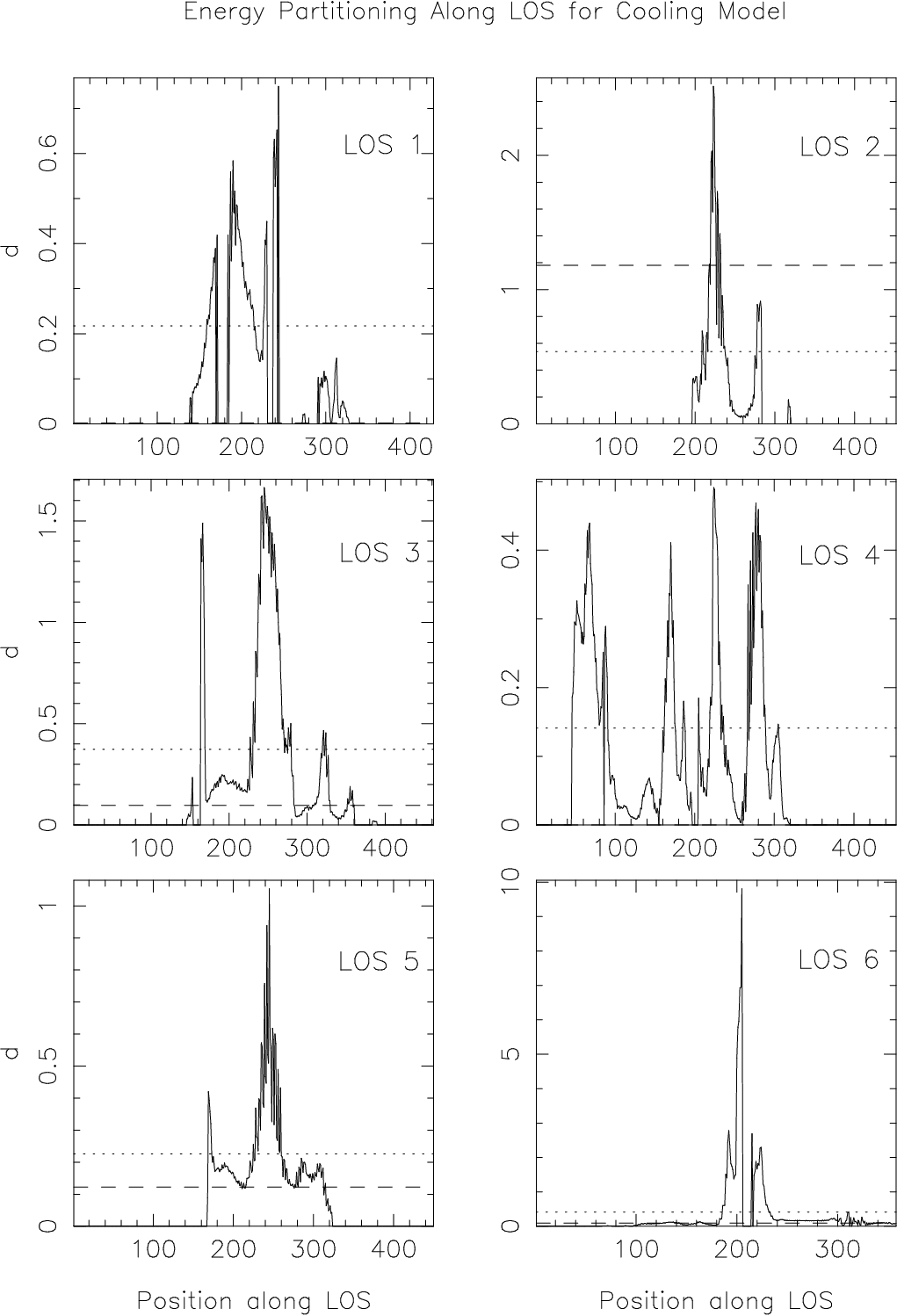}
\figcaption[f14.eps]
{Same as Figure \ref{dlos-m8.f} for the \cooling.
For reference, on axis the in-flowing jet carries $d_0 = 0.16$. 
\label{dlos-m7.f}}
\end{center}
\end{figure}

We again use the selected {\bf LOS 1~-~6} to 
explore how energy partitioning inferred from observations
compares to actual physical conditions within the sources.
Table \ref{dtable.t} lists the RMS and mean values, $d_{rms}$ and $\left< d \right>$,
along each line of sight, as well as the value of $d_{obs}$
inferred from the ratio $B_{ic}/B_{me}$ used in equation (\ref{dbme2.e}),
with $k = \eta = \sin{\vartheta} = 1$.
The direct values of $d$ were computed from the magnetic field
strength in each numerical bin and the nonthermal electron
energy density integrated between $p_0 = 10$ and 
$p_N = 1.63\times 10^5$,
corresponding to the synchrotron cutoffs assumed in calculating
$B_{me}$. 
Figure \ref{epart.f} shows graphically the relationship
between the $d_{obs}$ and $\left< d \right>$ data in Table \ref{dtable.t}.
For comparison, Figures \ref{dlos-m8.f}-\ref{dlos-m7.f} display distributions of
$d$ along {\bf LOS 1-6} using the actual, full nonthermal electron
distribution and the magnetic field properties from the 
simulations. 
Average values, $\left<d\right>$, are
indicated in those plots by the dotted lines and
$d_{obs}$ by dashed lines.

First looking at Figure \ref{epart.f} we see a rough correspondence
between $d_{obs}$ and $\left< d \right>$ for most of the data. On the other
hand agreement is not generally better than about an order of magnitude.
From equation (\ref{utot.e}) that would correspond to an uncertainty in the 
total energy 
of about a factor of four, if $U_{me}$ were known precisely.
We should not expect exact correspondence, since $d_{obs}$ is
based on two indirect measures weighted to different points
along each line of sight. We note below a couple of evident patterns
that are useful to examine in order to obtain better insights to
the limits of information contained in $d_{obs}$.

The two lobe LOS ({\bf LOS 1} and {\bf 4}) in the \cooling~ give $d_{obs}$
values two and four orders of magnitude, respectively, below $\left< d \right>$.
This strong bias is once again due to the influence of
spectral aging in the lobes of this model, which have not
been corrected for in this simple experiment. That is, if the
electron spectrum is convex, one must be careful to
count properly the low energy electrons, or $B_{me}$ will
be seriously over estimated, which has an even larger
impact on $d_{obs}$. Most of the other LOS
in the \control~ and the \cooling~ scatter reasonably around the
$d_{obs} = \left< d \right>$ line within about a factor of 4,
 and examination of Figures \ref{dlos-m8.f}
and \ref{dlos-m7.f} shows that $d_{obs}$ is usually ``representative''.
In that regard it is worth noting along the jet {\bf LOS 3}, {\bf 5}
and {\bf 6} that $d_{obs}$ agrees
well with physical values of $d$ in the jet itself, which
is the dominant emission source. On the other hand $d_{obs}$ for the
\control~ hotspot ({\bf LOS 2}) misses the actual energy
partitioning in the hotspot by about a factor of seven. Here
the field geometry is more complex, so that $B_{me}$ is an
under estimate of $B$ in the hotspot, even though it exceeds $B_{rms}$
along the LOS.

All of the $d_{obs}$ values in the \injection~
fall significantly below $\left< d \right>$. The closest
match comes from {\bf LOS 6}, which is dominated by the
(simple) jet base. Otherwise, in this model there are
very large fluctuations in $d$ that do not correspond
well to the strongest emission regions. So, in this
model $\left< d \right>$ is biased upwards compared to values
representative of the emitting plasma and that determine $d_{obs}$.  

The estimation of the local
energy partitioning $d$ based on $B_{ic}$ and $B_{me}$ works best in regions
that are not strongly influenced by radiative aging and where the partitioning
is not expected to vary strongly over short distances, such as inside jet 
structures.  There is increasing evidence that the X-ray emission observed
from some jets is doppler-boosted IC/3K emission 
\citep[\eg][]{Tavecchio00,Sambruna01,Celotti01}.
In that case this kind of analysis could be useful still, but it would require 
taking beaming and boosting effects into account, which we have not done 
because the flows here are nonrelativistic.  That obviously adds another
level of uncertainty, since those corrections can be large. Our analysis 
here was directed at the simpler case, but one that still contains
many real world features that must be accepted in any complex synchrotron
source.

\section{CONCLUSIONS} \label{bic:conc}

We have performed extensive synthetic radio and X-ray
observational analyses of the
numerically simulated radio galaxies introduced in TJR01.
These are the first synthetic observations with 
sufficient detail to allow the application of standard observational 
techniques to numerical simulations of radio galaxies.  
Standard observational
techniques were applied to the simulation data in order to understand
better how these techniques recover physical properties
of observed objects.  The simulated objects have the advantage
of known physical properties and evolve with many fewer simplifying
assumptions than required in analytic studies of this kind.
We emphasize that our goal was to compare observed properties
with known physical properties, not to present the simulated
objects as direct models for real objects. Our models
were intentionally idealized to isolate important nonthermal
particle transport behaviors.
We concentrated in this paper on magnetic field strength and
source energy content
calculations derived from synchrotron and inverse-Compton
intensities. An analysis of information extracted
through polarimetry will be presented in a forthcoming companion paper.

We enumerate some of the practical messages from our
analysis:

\begin{enumerate}
\item
{The synchrotron to inverse-Compton intensity ratio provides a
reasonable tool for estimating magnetic field strengths in
complex radio sources when the inverse-Compton
photons come predominantly from the CMB.
In our synthetically-observed 
simulated radio galaxies the standard radio/X-ray analysis returned 
magnetic field values that fell within about a factor two of the RMS field 
in a sampled volume, unless the electron spectrum was
strongly convex; \ie in the absence of strong aging. Strongly
aged spectra return field values that are too high, so need
to be corrected for that effect.
The effectiveness of this tool seems to apply even in regions where the 
electron distribution and the field structure are spatially intermittent;
\ie where they have small filling factors, such as in
the lobes of the simulated radio galaxies.  It is largely 
independent of the relative partitioning of energy
between electrons and magnetic field.}
\item
{The standard synchrotron emission minimum energy analysis returns 
magnetic field values consistent with the RMS fields only when the actual
energy partitioning between electrons and fields is close to equipartition.
Otherwise the inferred fields are biased in directions that
correctly reflect the actual deviations from equipartition in
the regions being sampled. In addition, our analysis
demonstrated that minimum energy estimates based
on strongly aged spectra can seriously overestimate
the actual mimimum energy magnetic field, unless the spectral
curvature is properly accounted for.}
\item
{The energy partitioning and the total source energy can be roughly 
estimated utilizing the
ratio of the minimum energy magnetic field 
to that inferred from the relative synchrotron and inverse-Compton
intensities. In our analysis the actual energy content was
recovered to within a little better than an order of magnitude,
once again in the absence of strong spectral aging. Without correction,
however, total energy contents were overestimated
by much more than an order of magnitude in regions with emissions 
dominated by strongly aged electron spectra.}
\item
{If sources are well resolved, it may be practical to examine
correlations between the radio and X-ray intensity distributions
as a probe of dynamical relationships between the particles
and the magnetic field. In our synthetic observation
analysis of these distributions
in the simulated objects we were able to recover correctly the 
physical correlation between magnetic field and plasma
density present in the objects.}
\item
{In regions with very large intensity contrasts, such as near a
bright hotspot, smoothing at low resolution naturally leads to
biases of the inferred properties in the direction of those
physically in the dominant emission regions.}
\end{enumerate}

Finally, a note on relativistic effects.  
Our surface brightness computations
would have to be modified in order to obtain
meaningful numbers for the case of relativistic flows.   However, our main purpose
has not been to study the details of the surface brightness distributions so much as
it has been to try to understand how well we recover meaningful information 
from the observations if we model them in simple but appropriate ways.  It
is difficult and probably of only limited usefulness to attempt a direct 
comparison to the 
relativistic case without 
genuine calculations.  A valuable extension 
of this work would be to apply the same kind of
analysis to appropriately-modeled emission (\ie including beaming and boosting
effects) from relativistic flows.  

\acknowledgements 
The work by I. L. T. and T. W. J. was supported by the NSF under grants 
AST96-16964 and AST00-71176 and by the University of Minnesota Supercomputing
Institute.  The work by D.R. was supported in part by KOSEF through grant 
R01-1999-00023.  We gratefully acknowledge Larry Rudnick 
for many helpful comments and discussions, the assistance of William
Ryan in developing data analysis tools, and the anonymous referee for
numerous valuable suggestions.  The Los Alamos National Laboratory strongly
supports academic freedom and a researcher's right to publish; therefore, the
Laboratory as an institution does not endorse the viewpoint of a publication
or guarantee its technical correctness.  

\appendix
\section{CALCULATION OF OMNIDIRECTIONAL SYNCHROTRON FLUX} \label{app:ssc}

Assuming an isotropic emissivity, 
$j_{s,\nu} \equiv j_{s}(\nu)$, 
and an optically thin
intervening plasma, the omnidirectional synchrotron flux at a given
position, $\Phi^{S}_{\nu}({\bf x})$, is estimated as
\begin{equation}
\Phi^{S}_{\nu}({\bf x}) = \frac{1}{4\pi} \int \frac{j_{s,\nu}({\bf x'})}
{({\bf x}-{\bf x'})^2} d^3{\bf x'}.
\label{app1.e}
\end{equation}
On a uniform grid of zone size $\Delta$,
the above equation is written as
\begin{equation}
\Phi^{S}_{\nu}(l,m,n) = \frac{\Delta}{4\pi} \sum_{l'=0}^{L-1}
\sum_{m'=0}^{M-1}\sum_{n'=0}^{N-1} G(l-l',m-m',n-n') j_{s,\nu}(l',m',n'),
\label{app2.e}
\end{equation}
where
\begin{equation}
G(l-l',m-m',n-n') \equiv \frac{1}{(l-l')^2+(m-m')^2+(n-n')^2}.
\label{app3.e}
\end{equation}
Here, $L$, $M$, and $N$ are the numbers of grid zones along the $x$, $y$,
and $z$-directions, respectively. Then, from the convolution theorem,
the omnidirectional flux can be computed as
\begin{equation}
\Phi^{S}_{\nu}(l,m,n) = \frac{\Delta}{4\pi} \mathcal{F}^{-1}
\left[\hat{G}^k(p,q,r)~\hat{j}_{s,\nu}^k(p,q,r)\right],
\label{app4.e}
\end{equation}
where $\hat{G}^k(p,q,r)$ and $\hat{j}_{s,\nu}^k(p,q,r)$ are the
Fourier transform of $G(l,m,n)$ and $j_{s,\nu}(l,m,n)$, respectively,
and $\mathcal{F}^{-1}$ denotes the inverse Fourier transform.
For $\hat{G}^k(p,q,r)$, we use
\begin{equation}
\hat{G}^k(p,q,r) = \frac{2\pi^2}{\left[\sin^2(2\pi p/L)
+ \sin^2(2\pi q/M) + \sin^2(2\pi r/N)\right]^{1/2}}.
\label{app5.e}
\end{equation}

We note that the discrete Fourier transform can be used only with
a periodic grid, while ours is not. However, as described, for instance,
in 
\citet{Binney87},
we can make $G(l,m,n)$ and
$j_{s,\nu}(l,m,n)$ periodic by doubling our grid along each direction
(that is, by extending the range of the summation into $-L$, $-M$, and
$-N$ in equation \ref{app2.e}). However, since the synchrotron emissions are generally
confined well inside the computational grid, aliasing is not significant.
We have established empirically that there was little difference between using
the doubled grid and using the original grid with periodic extensions.

\clearpage



\clearpage
\begin{table}
\caption{Summary of Simulations \label{summary.t}}

\begin{center}
\begin{tabular}{cccccc}

\tableline
\tableline
Model\tablenotemark{a} &
ID &
In-flowing &
Shock Injection  &
Cooling &
$B_{x0}$ \\
& & Electrons\tablenotemark{b}~($b_{1}$)& Parameter~($\epsilon$) &
Time\tablenotemark{c}~($Myr$) & ($\mu G$) \\

\tableline
1....... & Control & $10^{-4}$ & $0.0$ & $1.63 \times 10^{4}$ & $0.39$\\
2....... & Injection & $10^{-8}$ & $10^{-4}$ & $1.63 \times 10^{4}$ & $0.39$\\
3....... & Cooling & $10^{-4}$ & $0.0$ & $54$ & $5.7$\\
\tableline
\end{tabular}
\end{center}

\tablenotetext{a}{All models used external Mach $80$ jets
($M_{j} = u_{j}/c_{a} = 80$),
corresponding to a velocity of $0.05$ c, and a density contrast
$\eta = \rho_{j}/\rho_{a} = 0.01$; the internal jet Mach number is $8$.
Units derive from  $r_{j}=1$
(representing 2 kpc), an ambient density, $\rho_{a}=1$, and
a background sound speed, $c_{a}=(\gamma P_{a}/\rho_{a})^{1/2}=1~(\gamma=5/3)$.
The initial axial magnetic field was $B_{x0}$ ($\beta = P_{a}/P_{b} =
100$) in the ambient medium.  The jet also carried an additional toroidal
field component, $B_{\phi} = 2 \times B_{x0}(r/r_{j})$.  The spectrum of
nonthermal electrons entering with the jet was a power-law with momentum slope
$q=4.4$, corresponding to a synchrotron spectral index $\alpha = 0.7$.
The nonthermal particle distribution was specified by $N = 8$ momentum bins in
all three models.}

\tablenotetext{b}{Ratio of nonthermal to thermal electron density in the
incident jet flow.}

\tablenotetext{c}{Time for electrons to cool below momentum
$\hat p= 10^{4}m_{e}c$ in the background magnetic field $B_{x0}$.  In these
simulations the time unit $r_{j}/c_{a}$ corresponds in physical units
to approximately $10$ Myr.}

\end{table}

\clearpage
\begin{table}

\caption{$B_{ic}$ and $B_{me}$ Along LOS 1-6
\label{bic1.t}}
 
\begin{center}

\begin{tabular}{cccccccccccccc}
\tableline
\tableline
 & &
\multicolumn{6}{l}{Control Model ({\bf 1})~~~ LOS} \\
\cline{1-8} 

Field\tablenotemark{a} & Beam\tablenotemark{b} &
1 & 2 & 3 & 4 & 5 & 6 \\
\tableline
$B_{peak}$ & \nodata &
0.56 & 1.7 & 0.88 & 0.49 & 0.82 & 0.79 \\
$B_{rms}$  & \nodata & 
0.24 & 0.45 & 0.30 & 0.17 & 0.35 & 0.30 \\
\tableline
$B_{ic}$ \tablenotemark{d}
& 0.28 \tablenotemark{c} &
0.15  & 1.1  & 0.39 & 0.094 & 0.43 & 0.31 \\  


& 3.0  & 
0.13  & 0.91 & 0.40 & 0.088 & 0.43 & 0.31 \\



& 22.0 & 
0.18 & 0.37 & 0.34 & 0.19 & 0.36 &  0.29 \\
\tableline
$B_{me}$ \tablenotemark{e}
& 0.28 &
0.37  & 1.1  & 0.64  & 0.20  & 0.68 &  0.58 \\


& 3.0 &
0.30  & 0.91 & 0.65 & 0.19  & 0.70 &  0.58 \\



& 22.0 &
0.31 & 0.45 & 0.45 & 0.36  & 0.46 & 0.44 \\
\tableline
\tableline

& & \multicolumn{6}{l}{Injection Model ({\bf 2})~~~~ LOS} \\
\cline{1-8} 

Field\tablenotemark{a} & Beam\tablenotemark{b} &
1 & 2 & 3 & 4 & 5 & 6 \\
\tableline
$B_{peak}$ & \nodata &
0.56 & 1.7 & 0.88 & 0.49 & 0.82 & 0.79 \\  
$B_{rms}$  & \nodata & 
0.24 & 0.45 & 0.30 & 0.17 & 0.35 & 0.30 \\
\tableline
$B_{ic}$ \tablenotemark{d}
& 0.28 \tablenotemark{c} &
0.068 & 0.65 & 0.23 & 0.077 & 0.15 & 0.30 \\   

& 3.0  & 
0.052 & 0.69 & 0.89 & 0.087 & 1.5  & 0.29 \\

& 22.0 & 
0.95 & 0.89 & 1.2  & 0.15 & 1.1  &  0.16 \\
\tableline
$B_{me}$ \tablenotemark{e}
& 0.28 &
0.044 & 0.42 & 0.077 & 0.020 & 0.050 & 0.044 \\

& 3.0 &
0.049 & 0.39 & 0.11 & 0.023 & 0.39 &  0.060\\

& 22.0 &
0.13 & 0.23 & 0.23 & 0.037 & 0.24 & 0.13 \\
\tableline
\tableline

& & \multicolumn{6}{l}{Cooling Model ({\bf 3})~~~~ LOS} \\
\cline{1-8} 

Field\tablenotemark{a} & Beam\tablenotemark{b} &
1 & 2 & 3 & 4 & 5 & 6 \\
\tableline
$B_{peak}$ & \nodata &
8.3  & 25.3 & 13.1 & 7.3 & 12.1 & 11.7 \\
$B_{rms}$  & \nodata &
3.5  & 6.7  & 4.5  & 2.6  & 5.3  & 4.5 \\
\tableline
$B_{ic}$ \tablenotemark{d}
& 0.28 \tablenotemark{c} 
& 2.3   & 22.6 & 6.5  &  7.9 & 8.1 & 4.6 \\    


& 3.0   
& 2.8  & 17.3 & 6.5 & 5.5  & 8.0 & 4.8\\ 



& 22.0  
& 4.0 & 7.0 & 5.6 & 8.3 & 6.2 & 2.4 \\  
\tableline
$B_{me}$ \tablenotemark{e}
& 0.28 
& 15.3  & 24.2 & 14.2  & 242  & 16.6 & 10.2\\ 


& 3.0 
& 12.2  & 17.3 & 11.0 & 29.8  & 12.3 & 8.8 \\ 



& 22.0 
& 5.2  & 8.5  & 7.7  & 5.5   & 8.1  & 4.4 \\ 
\tableline

\end{tabular}

\end{center}
\tablenotetext{a}{All magnetic field values listed in \mug.}
\tablenotetext{b}{FWHM of convolved Gaussian in arcseconds.}
\tablenotetext{c}{Unconvolved pixel size in arcseconds.}
\tablenotetext{d}{Calculated from radio observation at 2.9 GHz and X-ray
observation at 1.2 keV.}
\tablenotetext{e}{Calculated using commonly-assumed $k=\eta=\sin{\vartheta}=1$.
Upper and lower cutoff frequencies are known from the model
parameters.  For Models {\bf1} and {\bf 2}, $\nu_1$ = 100 Hz and $\nu_2$ = 30
GHz. For Model {\bf 3}, $\nu_1$ = 1500 Hz and $\nu_2$ = 450 GHz. 
Calculation is based on synthetic radio surface brightness map
at 2.9GHz.}
\end{table}

\clearpage
\begin{table}
\caption{Energy Partitioning Along LOS 1-6 
\label{dtable.t}}
\begin{center}

\begin{tabular}{cccccc}  
\tableline
\tableline
Model &
ID &
LOS &
$d_{rms}$ &
$\left< d \right>$ &
$d_{obs}=1.5\left(\frac{B_{ic}}{B_{me}}\right)^{7/2}$ \tablenotemark{a}\\

\tableline
1 & Control\tablenotemark{b}    & 1 & 0.351  & 0.280 & 6.36 $\times 10^{-2}$ \\
  &           & 2 & 0.213  & 0.173 & 1.50   \\
  &           & 3 & 0.311  & 0.239 & 0.265  \\
  &           & 4 & 0.756  & 0.432 & 0.107  \\
  &           & 5 & 0.268  & 0.247 & 0.302  \\
  &           & 6 & 0.797  & 0.364 & 0.167  \\
\tableline
2 & Injection\tablenotemark{c} & 1 & 148    & 96.2 & 6.88  \\
  &           & 2 & 285    & 129  & 6.92  \\
  &           & 3 & 1110   & 716  & 69.1  \\
  &           & 4 & 821    & 479  & 168  \\
  &           & 5 & 1070   & 707  & 70.1  \\
  &           & 6 & 4340   & 1970 & 1240  \\
\tableline
3 & Cooling\tablenotemark{b}    & 1 & 0.278  & 0.217 & 1.98 $\times 10^{-3}$  \\
  &           & 2 & 0.783  & 0.539 & 1.18  \\
  &           & 3 & 0.584  & 0.373 & 9.73 $\times 10^{-2}$ \\
  &           & 4 & 0.192  & 0.141 & 9.43 $\times 10^{-6}$  \\
  &           & 5 & 0.276  & 0.226 & 0.122  \\
  &           & 6 & 1.170  & 0.413 & 9.24 $\times 10^{-2}$ \\
\tableline
\end{tabular}
\end{center}

\tablenotetext{a}{$d_{obs}$ found from equation (\ref{dbme2.e})
with $B=B_{ic}$ and $k = 1$.}
\tablenotetext{b}{Inflow boundary conditions set $d_0 = 0.16$ in the jet for these models.}
\tablenotetext{c}{Inflow boundary conditions set $d_0 = 1600$ in the jet for this model.}

\end{table}


\end{document}